\documentclass[12pt,preprint]{aastex}
\shorttitle{Cosmic rays and anisotropic transport}

\def\bm#1{\mbox{\boldmath $#1$}} 
\begin{document}
\title{Numerical simulations of buoyancy instabilities in galaxy cluster plasmas with cosmic rays and anisotropic thermal conduction}

\author{Y. Rasera and B. Chandran}
\affil{Space Science Center, University of New Hampshire, Durham, NH 03824, USA}
\email{yann.rasera@obspm.fr}

\begin{abstract}

In clusters of galaxies, the specific entropy of intracluster plasma
increases outwards.  Nevertheless, a number of recent studies have
shown that the intracluster medium is subject to buoyancy instabilities
due to the effects of cosmic rays and anisotropic thermal
conduction. In this paper, we present a new numerical algorithm for
simulating such instabilities. This numerical method treats the cosmic
rays as a fluid, accounts for the diffusion of heat and cosmic rays
along magnetic field lines, and enforces the condition that the
temperature and cosmic-ray pressure remain positive.  We carry out
several tests to ensure the accuracy of the code, including the
detailed matching of analytic results for the eigenfunctions and
growth rates of linear buoyancy instabilities.  This numerical scheme
will be useful for simulating convection driven by cosmic-ray buoyancy
in galaxy cluster plasmas and may also be useful for other
applications, including fusion plasmas, the interstellar medium, and
supernovae remnants.

\end{abstract}


\section{Introduction}

The hierarchical model of galaxy formation succesfully predicts the
evolution of baryons in the universe over a wide range of scales
assuming that supernov{\ae} feedback is taken into account
\citep{kauffmann99,somerville99,cole00,hatton03,springel03,rasera06}. However,
the baryon budget remains inaccurate in large scale structures (large
galaxies, groups or clusters) where the total amount of cold gas and
stars is overestimated. This overcooling problem is particularly
critical for galaxy clusters, in which the
cooling time near the center of a cluster is often much shorter
than a cluster's age. In the
absence of heating, one would expect cooling flows to form in these
clusters, with large amounts of plasma cooling and flowing in towards the
center. However, the star formation rate in cluster cores is typically
10-100 times lower than the predictions of the cooling-flow model
\citep{mcnamara04}, and line emission from plasma at temperatures
lower than one third of the virial temperature of the cluster is weak
\citep{peterson03,mcnamara04}. The inconsistency between the
cooling-flow model and these observations is known as the
``cooling-flow problem''.

A promising hypothesis to solve this puzzle is heating by active
galactic nuclei (AGN) in cluster cores. Two main arguments support
this idea. First, AGN power is expected to be a decreasing function of
the specific entropy at a cluster's center and therefore tends
naturally towards a self-regulated state in which heating balances
cooling \citep{nulsen04,boehringer04}. Second, almost all cooling core
clusters possess active central radio sources
\citep{eilek04}. However, one important problem remains: how is AGN
power transferred to the ambiant plasma? Over the last decade, a
number of numerical simulations have been carried out to answer this
question. In the first simulations
\citep{churazov01,quilis01,bruggen02a,bruggen02b}, thermal energy was
injected near the center of a 2D or 3D cluster-like hydrostatic
profile. This resulted in hot and underdense bubbles, which then rose
buoyantly. By agitating the surrounding medium, these bubbles were
able to reduce the cooling while achieving some correspondence with
the observations of X-ray cavities seen in roughly one-fourth of the
clusters of the Chandra archive \citep{birzan04}.  Subsequent
simulations extended these earlier works to include new physical
ingredients, such as viscosity
\citep{ruszkowski04a,ruszkowski04b,reynolds05,bruggen05,sijacki06}. It
was found that viscous dissipation contributed to the energy transfer,
and that viscosity helped to prevent bubbles from breaking up. Other
studies \citep{reynolds01,reynolds02, omma04a,omma04b, cattaneo06,
  heinz06} injected not only thermal energy but also kinetic energy in
subrelativistic bipolar jets. This approach also leads to cavities,
but the dynamics are different than in the previous works because of the initial
momentum of the bubbles and because the energy is deposited over a more narrow range of angles. In this context, the importance of
turbulence, magnetohydrodynamics effects, and plasma transport
processes has been underlined by \citet{vernaleo06}, who suggested that
these ingredients could prevent the heating from being highly
concentrated along the jet axis, as is the case for one-fluid
pure-hydrodynamics simulations of jets in clusters that are initially
at rest.

The above simulations treated the intracluster medium (ICM) as a single
fluid.  In single-fluid simulations, when AGN-heated plasma at temperature
$T_{\rm hot}$ mixes with ambient intracluster plasma at
temperature~$T_0$, the result is a Maxwellian plasma with a
temperature intermediate between $T_0$ and $T_{\rm hot}$.  Although
this approach is valid in clusters if~$T_{\rm hot}$ is not too large,
it breaks down if the hot particles are relativistic or
transrelativistic, because then Coulomb collisions do not have
sufficient time to bring the hot particles into thermal equilibrium
with the ambient intracluster plasma.  If we focus on hot protons, the
type of collision that brings such protons most rapidly into thermal
equilibrium with the background plasma is collisions with background
electrons.  The time scale for thermal electrons to remove energy from a
hot proton is \citet{gould72},
\begin{eqnarray}
\tau_\epsilon=\frac{(\gamma-1) m_p m_e v_p c^2}{4 \pi e^4 n_e} \left[\textrm{ln}\left( \frac{2 m_e c v_p p}{\hbar (4 \pi e^2 n_e/m_e)^{\frac{1}{2}}} \right)-\frac{v_p^2}{2 c^2}\right],
\label{eq:taue}
\end{eqnarray}
with $n_e$ the electron density, $\gamma$ the Lorentz factor, $v_p$ the proton velocity, $e$ and $m_e$ the electon charge and mass, $p$ and $m_p$ the proton momentum and mass, and $\hbar$ the reduced Plank constant.

For a typical proton energy of $E \simeq 1$~GeV (transrelativistic regime) and a typical cluster-core electrons density of $n_e=0.01$~cm$^{-3}$, the thermalization time scale is $\tau_\epsilon \simeq 7$~Gyr, which is much larger than the time for protons to
escape the cluster core via diffusion or convection. In this case,
the ICM is essentially a two-fluid system similar to the
interstellar medium of the galaxy, with a thermal background plasma
plus a population of high-energy particles (cosmic rays).

There are a few problems with treating a mix of cosmic rays and
thermal plasma as a single Maxwellian fluid. One is that the
single-fluid approximation to the temperature contains the cosmic-ray
contribution to the energy density, and thus overestimates the actual
temperature of the thermal plasma. If the cosmic-ray energy density is
a significant fraction of the total energy density, the single-fluid
model is unable to accurately predict the temperature profile of a
cluster. In addition, since the thermal conductivity depends sensitively on the
temperature ($\kappa_T \propto T^{5/2}$), and since conduction can
make an important contribution to the heating of a cluster core \citep{zakamska03}
errors in the temperature profile can also lead to significant secondary
errors in the thermal balance of the ICM.

A more subtle difficulty in applying a one-fluid model to a
cosmic-ray/thermal-plasma mixture concerns the convective stability of
intracluster plasma. It turns out that a radial gradient in the cosmic-ray energy density is much more destabilizing than a radial gradient in the thermal plasma energy density when the plasma mass density decreases outwards (see Eq.\ref{criterion} below). A correct accounting of the fraction of the total pressure contribution by cosmic rays is thus essential for understanding the convective stability of clusters. A more extensive discussion of this point is given by \citet{chandran06}.


A more accurate treatment of the ICM, in which the cosmic rays are
treated as either a second fluid or as collisionless particles, is
thus needed. In this paper, we present a new numerical algorithm for
simulating the ICM that treats the ICM as a two-fluid (cosmic-ray plus
thermal-plasma) system. We also present the results of a suite of
tests for our code. Our numerical approach is similar to that
of Mathews \& Brighenti (2008), who carried out two-fluid simulations
of cosmic-ray bubbles in the ICM. However, in contrast to this
latter study, we take thermal conduction and cosmic-ray diffusion to
occur almost entirely along magnetic field lines (cross-field
transport arising only from numerical diffusion). Such anisotropic
transport arises in clusters because the Coulomb mean free paths of
thermal particles in clusters are much larger than their gyroradii, and
the scattering mean free paths of cosmic rays are much larger than
their gyroradii. The effects of magnetic fields on conduction are some
times taken into account in when considering thermal conduction over
length scales much larger than the correlation length of the (tangled)
intracluster magnetic field, $l_B \simeq 1-10$~kpc
\citep{kronberg94,taylor01,taylor02,vogt03}.  In this case, the
conductivity~$\kappa_T$ is effectively isotropic
\citep{rechester78,chandran98} with a value that is $\simeq 0.1-0.2$
times the Spitzer thermal conductivity for a non-magnetized plasma
\citep{narayan01, chandran04b, maron04}.  However, on scales~$\lesssim
l_B$, the anisotropy of the thermal conductivity has a powerful effect
on the convective stability of the intracluster medium
\citep{balbus00, balbus01, parrish05,chandran06, parrish07, parrish08,
  quataert08}, in such a way as to make convection much more likely
than when the conductivity is treated as isotropic.  This is true even
if the magnetic field is so weak that the Lorentz force is negligible.
In order to simulate buoyancy instabilities and convection in clusters,
it is thus essential to incorporate anisotropic transport.

The remainder of this paper is organized as follows. In
section~\ref{sec:equations} we present the basic equations of our
two-fluid model. In section~\ref{sec:notransport} we present the
total-variation-diminishing (TVD) code that we use to solve these
equations as well as several numerical tests, focusing on the case in
which there is no conduction or diffusion. In section~\ref{TVD
  anisotropic transport} we present the standard numerical discretization
method for anisotropic conduction. We show how it can lead to negative
temperature as emphasized before by \citet{sharma07}. We then describe
our new method that does not suffer from negative temperature
problems. Tests such as the circular conduction test and Sovinec-test
are also presented. Finally, in section~\ref{sec:buoyancy} we present
results for the linear buoyancy instabilities involving cosmic rays
and anisotropic transport and compare our numerical solutions to
analytic results.


\section{Two-fluid equations with anisotropic transport}
\label{sec:equations} 

In order to carry-out realistic cluster simulations one has to deal
with an impressive list of components (dark matter, plasma, cosmic
rays, magnetic field, stars, supernov{\ae}, supermassive black holes)
and physical ingredients (advection, shocks, induction, gravity,
anisotropic transport, cooling, energy injection from the AGN,
feedback from supernov{\ae}, jets, viscosity...). In this paper, rather
than attempting to simulate all of these processes, we focus on developing an accurate and efficient 
numerical algorithm for simulating collisional plasmas pervaded by collisionless cosmic
rays. A complete description of the cosmic rays in such a system would require us to solve
for the cosmic-ray distribution function~$f(\bm{r},\bm{p}, t)$, where
$\bm{r}$ is the position coordinate, $\bm{p}$ is momentum, and~$t$ is
time. The resulting system of equations is much more difficult to
solve numerically than a system of fluid equations because $f$ depends on three
momentum coordinates as well as position and time.  However, in many
situations of interest, $f$ is nearly isotropic in momentum space and
can be treated as function of only $(\bm{r}, |\bm{p}|, t)$ \citep{skilling75}. 
\citet{miniati01alone,miniati01} took advantage of this
fact with a numerical code, COSMOCR, that solves for the evolution of
$f$ as a function of both $\bm{r}$ and $p$. In this paper, we adopt
the more simplistic and less computationally intensive fluid-like approach of
\citet{drury81}, which does not attempt to solve for the momentum
dependence of~$f$, but instead solves directly for the evolution of the
cosmic-ray pressure~$p_{\rm cr}$ as a function of~$\bm{r}$ and~$t$.
This model of \citet{drury81} has been extended to three spatial
dimensions by \citet{jones90}, to include gravity by \citet{mathews08}, 
and to include the magnetic field and
Lorentz force by \citet{ryu03}. In this paper, we extend the model further to
include anisotropic thermal conduction.  The resulting equations can be
written,

\begin{eqnarray}
\frac{\partial \rho}{\partial{t}}+\bm{\nabla}.(\rho \bm{v})&=&0\label{eqMass}\\ 
\frac{\partial \rho \bm{v}}{\partial{t}}+\bm{\nabla}.\left(\rho \bm{v}\bm{v}+p_{tot}-\frac{\bm{B}\bm{B}}{4\pi}\right)&=&\rho \bm{g} \label{eqImpuls}\\
\frac{\partial \bm{B}}{\partial{t}}-\bm{\nabla} \times (\bm{v} \times \bm{B})&=&0 \label{eqMag}\\
\frac{\partial e}{\partial{t}}+\bm{\nabla}.\left((e+p_{tot})\bm{v}-\frac{\bm{B}(\bm{B}.\bm{v})}{4\pi}\right)&=&\rho \bm{v}.\bm{g}+\bm{\nabla}.({{\bm\kappa}}.\bm{\nabla}T)+\bm{\nabla}.({{\bm D}}.\bm{\nabla}e_{cr})\label{eqNRJ}\\
\frac{\partial e_{cr}}{\partial{t}}+\bm{\nabla}.(e_{cr}\bm{v})&=&-p_{cr}\bm{\nabla}.\bm{v}+\bm{\nabla}.({{\bm D}}.\bm{\nabla}e_{cr})\label{eqCR},
\end{eqnarray}
with the 9 main variables, $\rho$ the plasma density, $\bm{\rho v}$
the plasma momentum, $\bm{B}$ the magnetic field, $e_{cr}$ the cosmic
ray internal energy, and $e=0.5 \rho v^2+e_{th}+e_{cr}+0.5B^2/4\pi$
the total energy. Intermediate variables are $e_{th}$, the internal
thermal energy and $p_{tot}=(\gamma-1) e_{th}+ (\gamma_{cr}-1)
e_{cr}+0.5B^2/4\pi$ the total pressure with $\gamma$ and $\gamma_{cr}$
the adiabatic indices of gas and cosmic rays. $\bm{g}$ is an external
gravity field (the large scale gravitational potential is mostly
dominated by stars and dark matter in galaxy clusters).  Finally,
${{\bm D}}$ and ${{\bm\kappa}}$ are the diffusion and conduction
tensor, which are described further in section~\ref{TVD anisotropic transport}.

In this model, the cosmic rays flow at the same speed as the thermal
plasma, since both are frozen to the same magnetic field lines and
since wave-particle interactions limit the relative motion between cosmic rays and
thermal plasma in the direction of the magnetic field. On the other
hand, because the pitch-angle scattering associated with wave-particle
interactions is of finite strength, the cosmic rays can diffuse with
respect to the thermal plasma. We have taken this diffusion, as well
as the conduction of heat, to occur entirely along magnetic field
lines. This is a reasonable approximation in clusters of galaxies,
because the gyroradii of thermal particles are much shorter than their
collisional mean free paths, and the gyroradii of cosmic rays are much
shorter than their scattering mean free paths. The value of the
cosmic-ray diffusion coefficient in clusters of galaxies is not well
known. In this paper, we use a value $D_\parallel = 10^{29} \mbox{
 cm}^2 \mbox{s}^{-1}$ comparable to the parallel diffusion coefficient of
1~GeV protons in the interstellar medium of our galaxy.

We assume that protons dominate the cosmic-ray energy density in
clusters, as is the case in the Galaxy. For protons in clusters, the
energy loss times associated with Coulomb interactions and inelastic
collisions (pion production) are typically longer than the growth
times of the instabilities that we focus on in this paper. Thus, we
neglect Coulomb losses and pion production in this paper. We note that
we also do not include self-gravity, since it is not important in the
hot intracluster medium.

It can be seen from equation \ref{eqCR} that the cosmic rays are treated as
a fluid with adiabatic index~$\gamma_{\rm cr}$. Thus, if $\nabla \cdot
\bm{v} < 0$ at some location, the converging flow acts to increase the
cosmic-ray pressure.  If we were to model the cosmic-ray distribution
function~$f(\bm{r},\bm{p},t)$ as a power law in momentum of the
form~$p^2 f \propto p^{-\alpha}$ with a low energy cutoff, then the
effective value of $\gamma_{\rm cr}$ is given by equation~[13] of
\citet{jubelgas08}.  In this paper we make the simple choice that
\begin{equation}
\gamma_{\rm cr} = 4/3,
\end{equation}
corresponding to the limit in which $\alpha$ approaches~2 from above,
and in which the cosmic-ray energy density is dominated by
ultra-relativistic particles.

We note that our approach is in some ways similar to the model of
\citet{jubelgas08}, who incorporated cosmic rays into
hydrodynamical simulations of galaxy formation based on smoothed
particle hydrodynamics.  Their approach, like ours, employs an
effective adiabatic index for the cosmic rays and avoids solving for
the full cosmic-ray momentum distribution function. However, \citet{jubelgas08}
 also develop a framework for incorporating a number of
effects that are not treated here, including ionization losses,
radiation losses, and shock acceleration.  On the other hand, \citet{jubelgas08} 
assume an isotropic cosmic-ray diffusion coefficient,
whereas anisotropic transport of both cosmic rays and heat plays a central
role in our model as well as the buoyancy instabilities that we
simulate in section~\ref{pcisection}.

\section{Numerical implementation and tests in the absence of transport}
\label{sec:notransport} 


In this section, we set $\bm{\kappa}=\bm{D}=\bm{0}$. Our numerical
method for solving the magnetohydro\-dynamic-like (MHD-like) two-fluid equations is based on the
Total Variation Diminishing (TVD) MHD code of \citet{pen03} which has the
advantage of being fast, simple and efficient. This TVD MHD code is
fully described in 3 papers. The appendix of \citet{pen98} presents
the relaxed TVD method that is used, \citet{trac03} shows the different methods for
hydrodynamics solver and \citet{pen03} describes the MHD code
itself. We will here recall the main characteristics of this code but
the reader should refer to the above articles for more complete
explanations.

The fluid solver is a conservative, second-order (in space and time),
dimensionally split, TVD, upwind scheme. In this relaxing TVD method,
each hyperbolic conservation law is replaced by a left and a right
advection problem with an advection speed called the ``freezing
speed''. By taking this freezing speed equals to the largest
eigenvalue of the system $c=max(|v|+c_s)$ (with $c_s$ the sound speed
and $v$ the velocity along the updated direction), it ensures the
scheme to be TVD. The advection problem is then solved using Van-Leer
slope limiter to reach second order in space and Runge-Kutta
integration to reach second order in time.

\begin{figure}
  \plotone{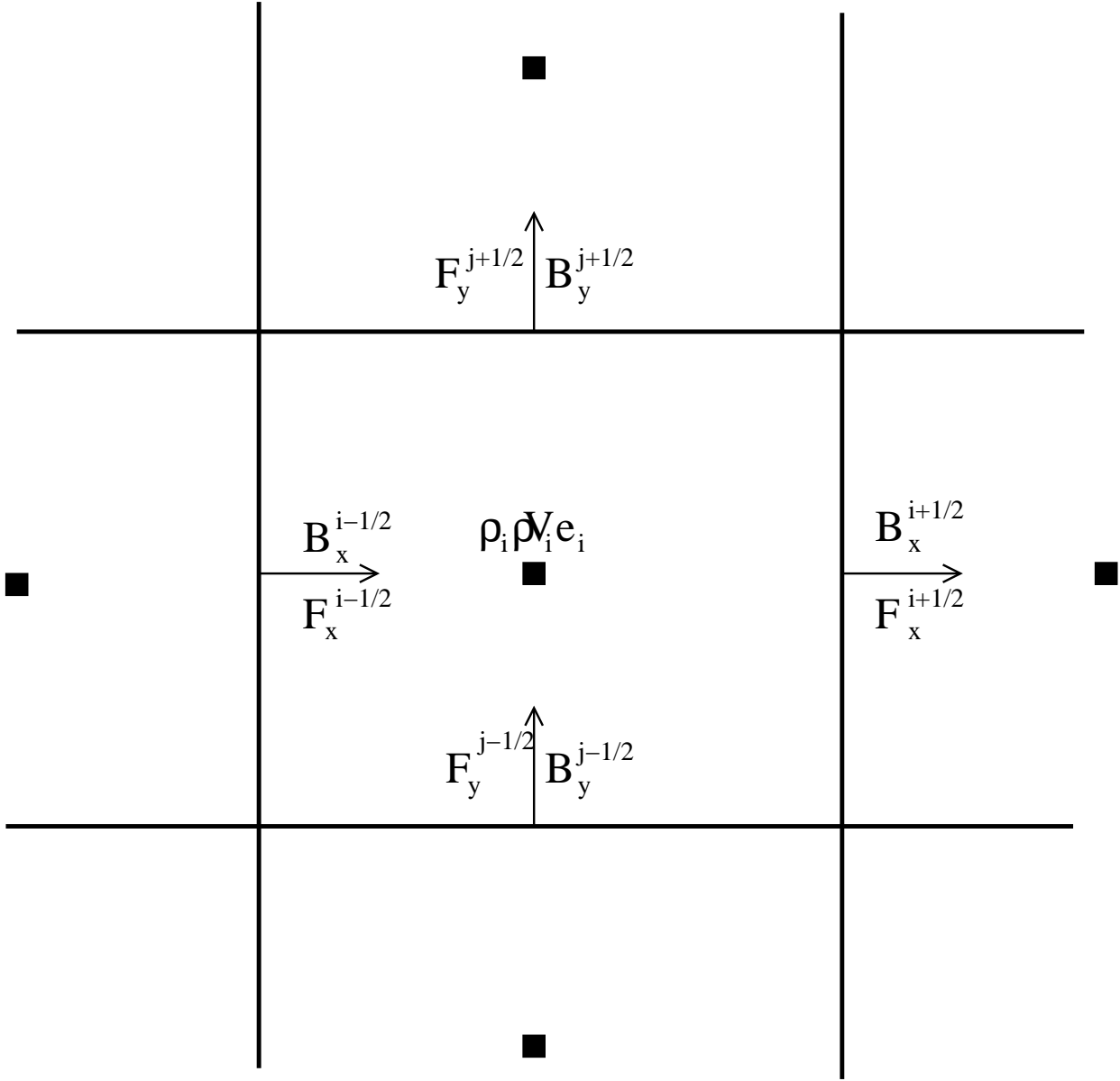}
  \caption{Staggered grid used by the MHD solver. Density, momentum and energy are defined at the cell center whereas magnetic fields are defined on the faces. Fluxes are also computed on the faces and depend on the neighbours (squares).\label{gridMHD}}
\end{figure}

The magnetic field is updated separately in advection-constraints
step. A staggered grid is used with $\bm{B}$ defined on cell surfaces
(see Fig.\ref{gridMHD}) in order to satisfy the divergence-free
magnetic field condition at machine precision. The advection step is
computed using the same TVD method as in the fluid solver. This is
however easier since the velocity is assumed to be fixed (operator
splitting). The second order flux is then directly re-used to compute
the constraint step. Here again Runge-Kutta is used for second order
temporal accuracy.

This method is very efficient because it doesn't need to solve the
whole Riemann problem and therefore compute all eigenvalues. It only
needs the computation of the largest one for the freezing speed. The
resolution of slow waves is slightly degraded, however the code could
still resolve shocks using only a few cells.

The fluid solver has succesfully been tested for advection of a square
wave and evolution of a three dimensionnal Sedov blast wave. Finally,
the MHD code gives good results on various tests such as slow, fast
and Alfv\`en waves as well as an MHD shock-tube problem.

To modify this TVD code to solve
equations~(\ref{eqMass}) through (\ref{eqCR}), we include an
additional fluid variable $e_{\rm cr}$ and  use exactly
the same routine. The flux vector associated with the conservative variables
becomes (for an update along 'x'),
\begin{displaymath}
\bm{F}=\left(  
\begin{array}{c}
\rho v_x\\
\rho v_x^2+p+0.5B^2/4\pi+p_{cr}-B_x^2/4\pi\\
\rho v_x v_y-B_x B_y/4\pi\\
\rho v_x v_z-B_x B_z/4\pi\\
(e_{tot}+p+0.5B^2/4\pi+p_{cr})v_x-B_x\bm{B}.\bm{v}\\
e_{cr} v_x\\
\end{array}\right),
\end{displaymath}
with $p_{cr}=(\gamma_{cr}-1)e_{cr}$. The freezing speed becomes $c=max[|v_x|+\sqrt{(\gamma p+\gamma_{cr} p_{cr}+ B^2/4\pi)/\rho}]$ and the timestep is reduced to $dt=0.8\Delta x/max[(|v_x|,|v_y|,|v_z|)+\sqrt{(\gamma p+ \gamma_{cr} p_{cr}+B^2/4\pi)/\rho}]$. In this way, we recover the original MHD TVD method when $p_{cr}=0$ and otherwise take into account effects of cosmic rays using the same TVD routine.

The remaining source term $-p_{cr}\bm{\nabla}.\bm{v}$ is related to the pressure work during expansion or contraction and is easy to implement. Following the general philosophy of the code, we discretize it at second order accuracy using dimensional splitting for multidimensional runs and Runge-Kutta to reach second order temporal accuracy. It leads to:
\begin{eqnarray}
-p_{cri}^n \frac{v^n_{i+1}-v^n_{i-1}}{2 \Delta x}.
\end{eqnarray}
This contribution is finally included in the energy update at the beginning of each one-dimensional hydrodynamics step. In the following sections, we describe several tests of this 2-fluids code that we have performed.

\subsection{Linear test: propagation of a sound wave in a composite of cosmic rays and thermal gas}

The first simple test is the propagation of sound wave in a medium with cosmic rays and plasma. The adiabatic wave speed is given by
\begin{eqnarray}
c_s=\sqrt{ \frac{\gamma p_0+\gamma_{cr} p_{cr0}} {\rho_0} },
\end{eqnarray}
with $\rho_0$, $p_0$ and $p_{cr0}$ the initial non-perturbed quantities.
In order to trigger an eigenfunction, we need to satisfy the following relations between the field perturbations,
\begin{eqnarray}
\frac{\delta \rho}{\rho_0}&=&\frac{\delta v}{c_{s}} \label{pertrho},\\
\frac{\delta p}{p_0}&=&\gamma \frac{\delta v}{c_{s}}\label{pertp},\\
\frac{\delta p_{cr}}{p_{cr0}}&=&\gamma_{cr} \frac{\delta v}{c_{s}}\label{pertpcr},
\end{eqnarray}
with $\delta v$, $\delta \rho$, $\delta p$ and $\delta p_{cr}$ the perturbations. For our test, we take for the equilibrium quantities $v_0=0$, $\rho_0=1$, $p_0=1$ and $p_{cr0}=1$. We then perturbate the velocity with a sine of amplitude $\delta v= 10^{-3}$ and wavelength $0.5$. The other quantities are perturbated following \ref{pertrho}, \ref{pertp} and \ref{pertpcr}.

\begin{figure}[h!]
  \plotone{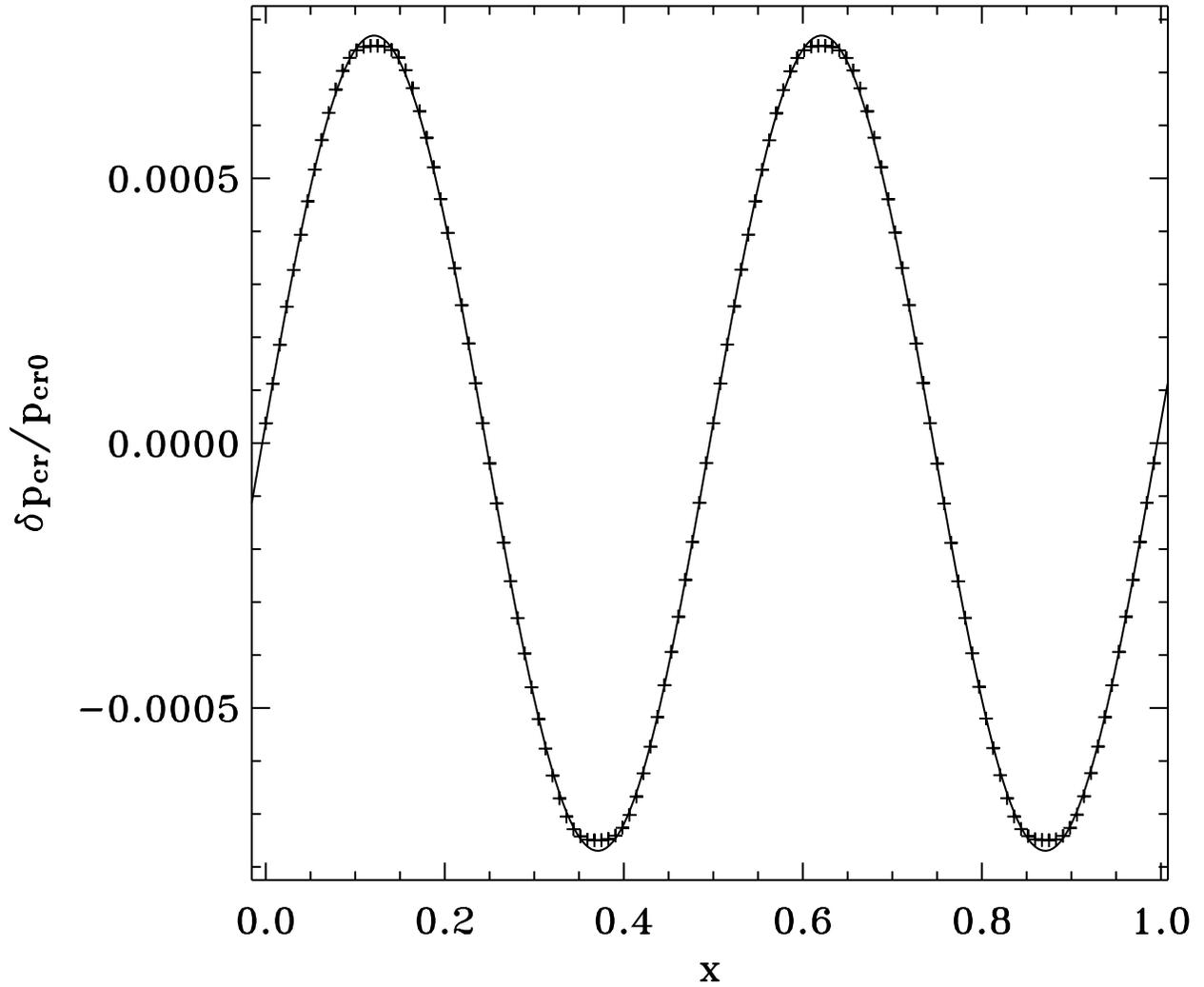}
    \caption{Sound wave in a composite of cosmic rays and thermal gas propagated for one wave period. Plus signs represent the simulation cosmic-ray internal energy for 128 grid points whereas the continuous line is the analytical solution.}
    \label{soundwave}
  \end{figure}

Fig.\ref{soundwave} shows the results after a propagation during one period for $128$ grid points. The result is in good agreement with the analytical solution and we obtain the same level of accuracy achieved in a pure hydrodynamical simulation (without cosmic rays). The slight smoothing of the extrema is due to the slope limiter which prevents the code from introducing spurious oscillations.

\subsection{Non Linear test: Riemann shock-tube problem for a composite of cosmic rays and thermal gas}

A more challenging test is the Riemann shock-tube problem. The standard problem \citep{sod78,hawley84} involves a polytropic gas starting with a state of high pressure and high density in the half-left space and a state of low density and low pressure in the half-right space. It leads to 5 regions with different fluid states separated by the head and tail of the rarefaction wave, the contact discontinuity and the shock. The interesting point is that one can derive the analytical solution using the Rankine-Hugoniot conditions.

However, in our case the composite of cosmic rays and thermal gas is not a polytropic fluid and this solution doesn't apply. This problem has been solved by \citet{pfrommer06} and we use here their analytical solution. Our 1D initial conditions are close to the ones used in their article with a left-hand state (L) and a right-hand state (R) in the simulation box. They are given (using an appropriate system of units) by,
\begin{eqnarray}
1<x<1.5  && 1.5\leq x<2,\\
\rho_L=1 && \rho_R=0.2,\\
v_L=0    &&  v_R=0,\\
p_L=6.7\times 10^4&&p_R=2.4\times 10^2,\\
p_{crL}=1.3\times 10^5&&p_{crR}=2.4\times 10^2.\\
\end{eqnarray}
The sound speed is therefore $c_{sL}=537$ and $c_{sR}=60$. We run a simulation with 1024 grid points until $t=4.4\times 10^{-4}$ so that the shock front has propagated on an important fraction of the box length $L=1$.

\begin{figure*}[h!]                                  \begin{tabular}{cc}
    \includegraphics[width=0.45\hsize]{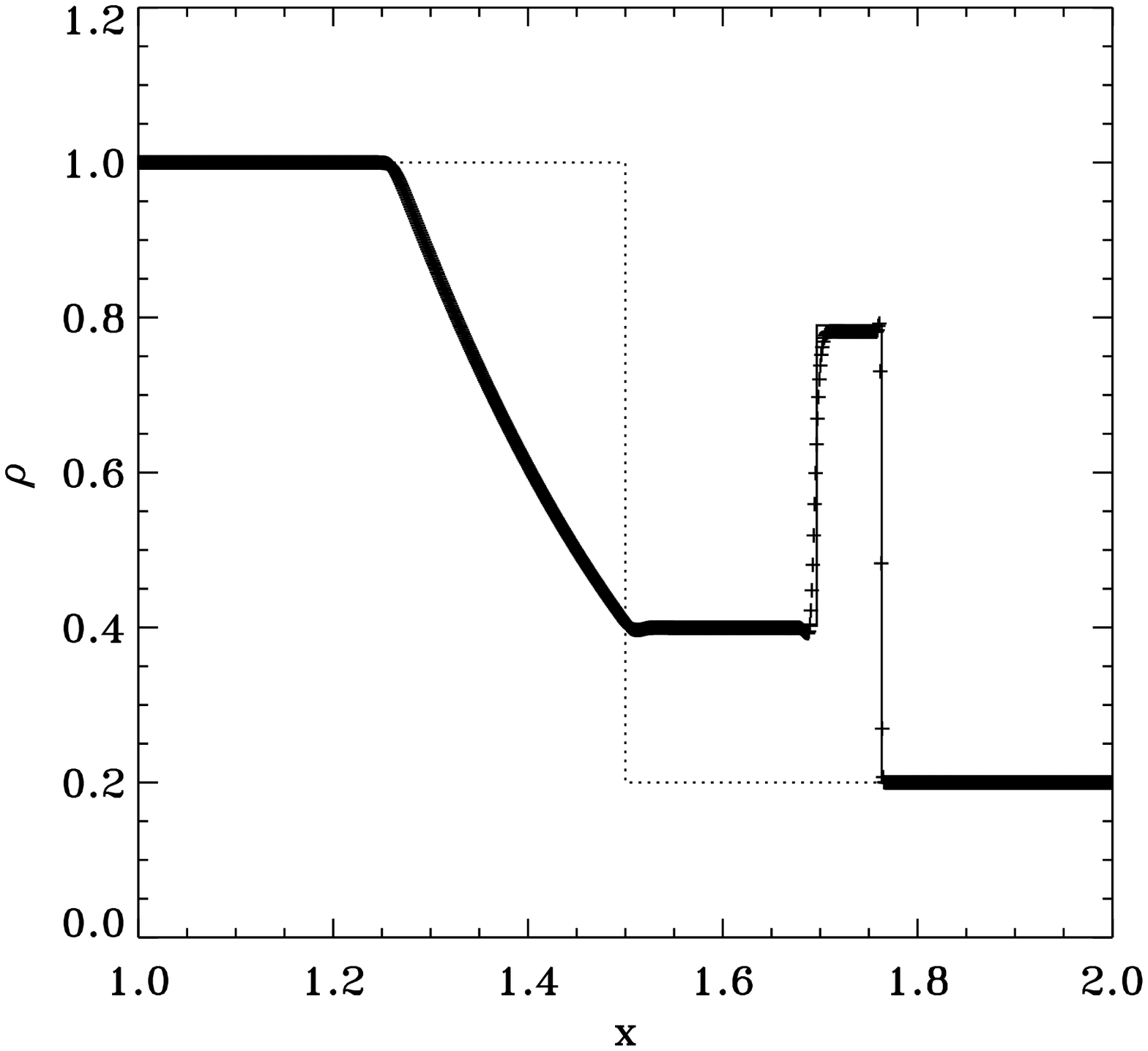}
    \includegraphics[width=0.45\hsize]{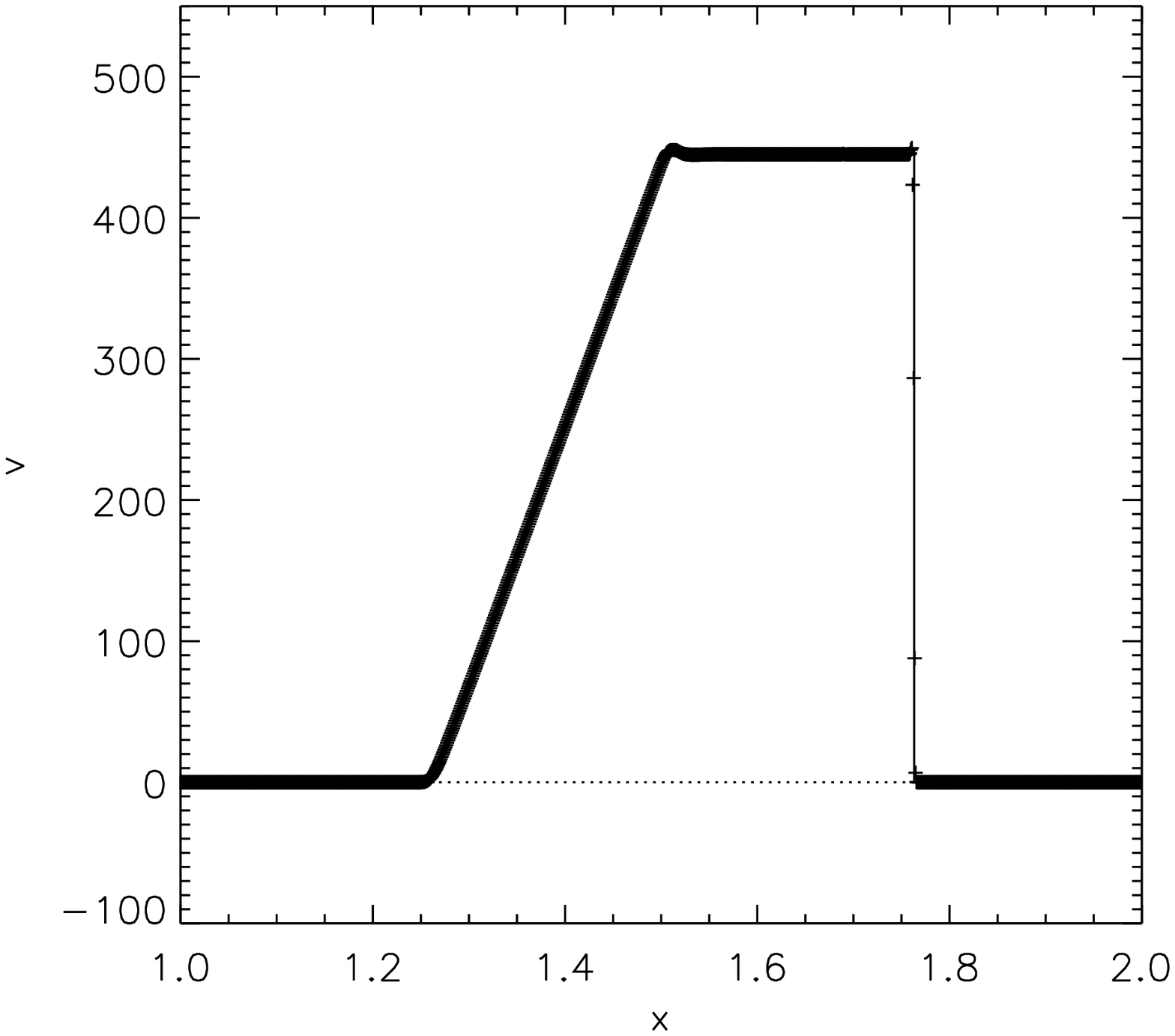}             \\
    \includegraphics[width=0.45\hsize]{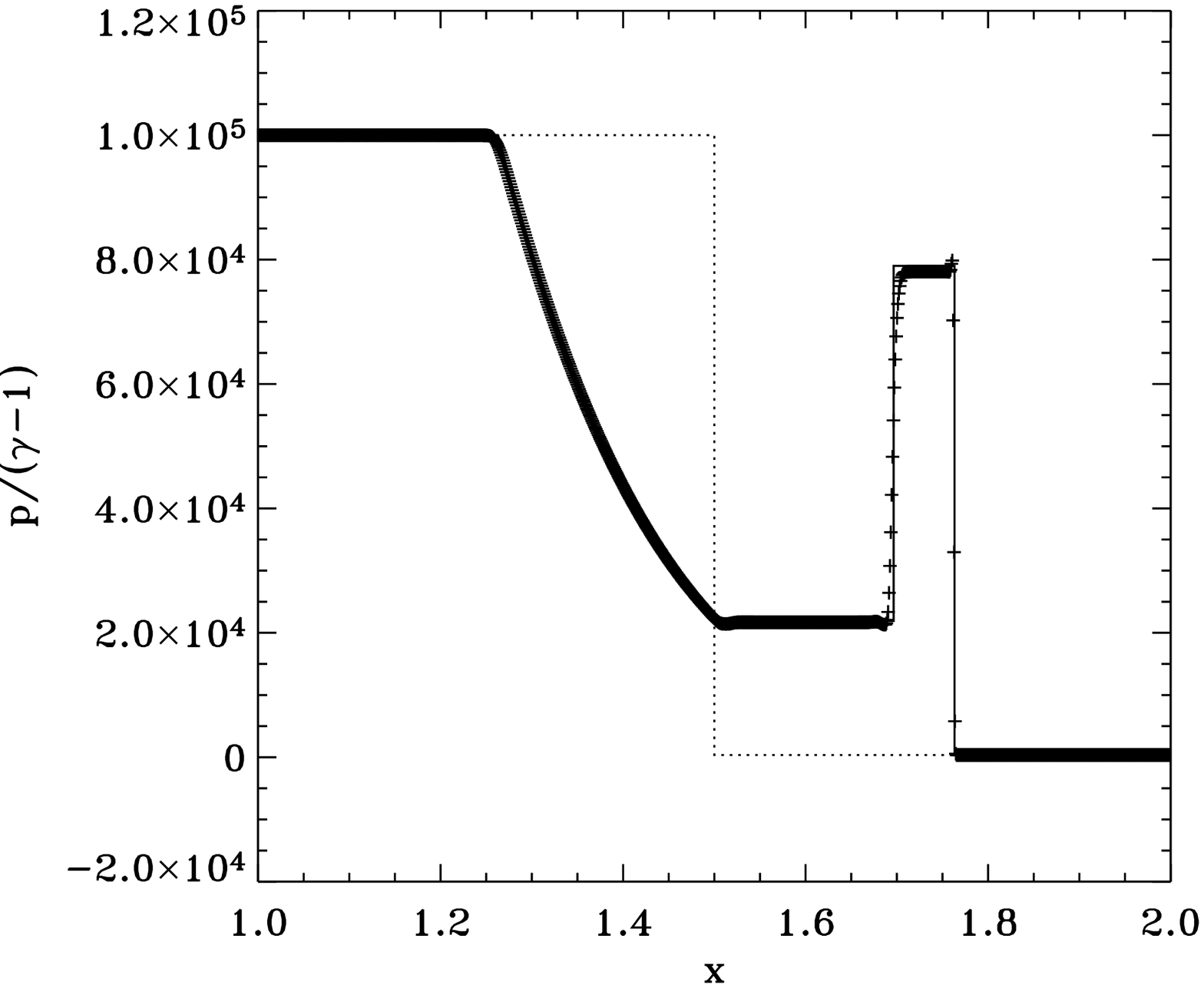}
    \includegraphics[width=0.45\hsize]{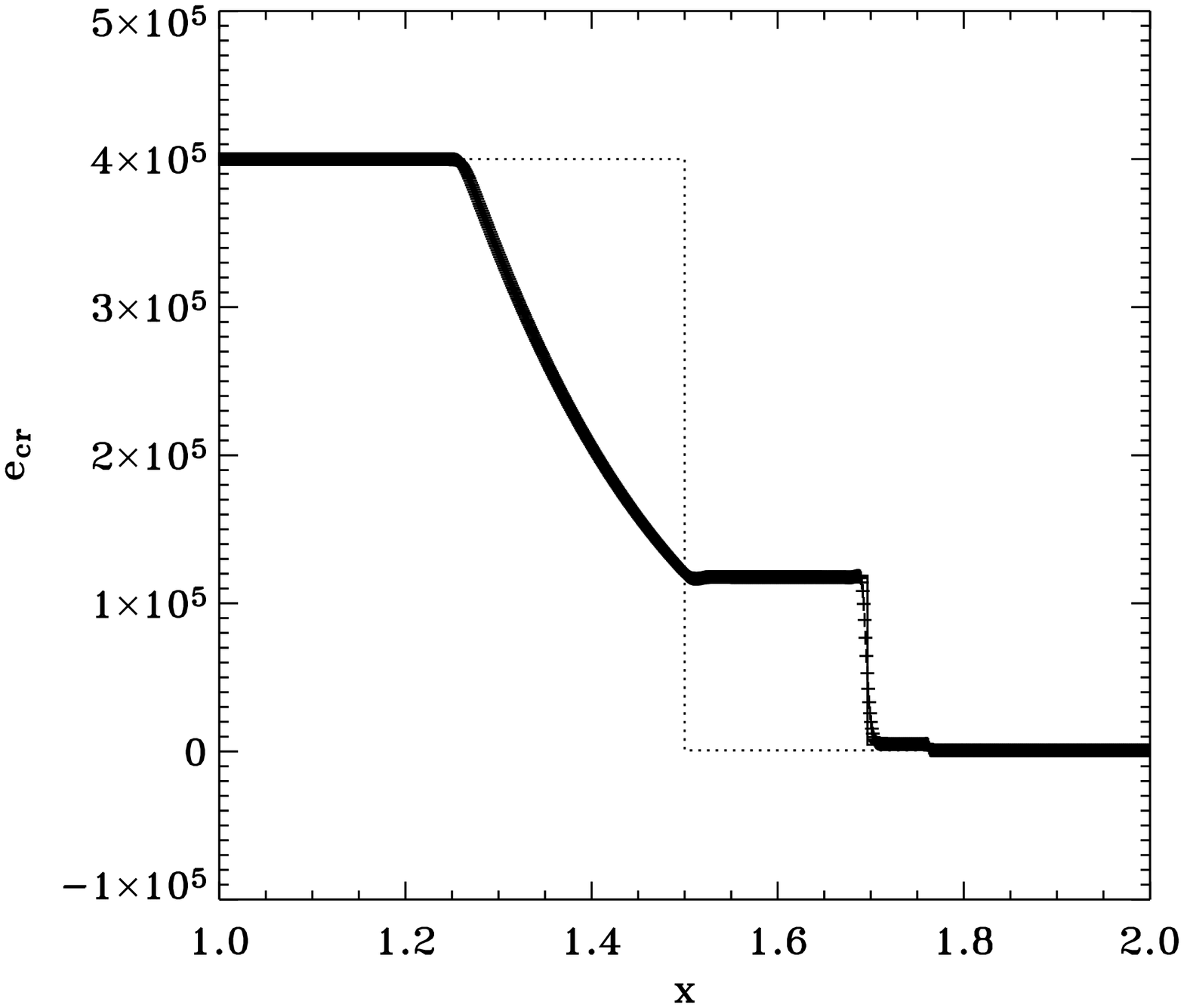}
    \end{tabular}  
  \caption{Shock tube problem for a composite of cosmic rays and thermal gas. Plus signs show results for a 1024 grid points simulation whereas the continuous line is the analytical solution. The dotted line is the initial condition. The first graph shows the density profile, the second one the velocity profile, the third one the gas internal energy profile and finally the fourth one shows the cosmic-ray internal energy.} \label{tubecr} 
\end{figure*}

Here again, there is a good agreement between the simulation results and the analytical prediction (see Fig.\ref{tubecr}). The transitions between the 5 states are well situated. The shock is resolved using few cells. As in \citet{pen03} some variables have a slight overshoot in the first postchock cell but this doesn't affect the other subsequent cells. The contact discontinuity is slightly smoothed by the relaxation solver. To conclude, the accuracy is similar to the accuracy obtained in the 1-fluid shock tube test and the implementation of cosmic rays is successful.

\section{Anisotropic transport: heat conduction and cosmic-ray diffusion}
\label{TVD anisotropic transport}


In the presence of a magnetic field, the heat conduction
in a plasma takes the form
\begin{eqnarray}
\frac{\partial e}{\partial t}=\bm{\nabla}.\left(\kappa_{\|}\hat{\bm{b}}\hat{\bm{b}}.\bm{\nabla}T\right)+\bm{\nabla}.\left[\kappa_{\bot}({{\bm I}}-\hat{\bm{b}}\hat{\bm{b}})\bm{\nabla}T\right] \label{heat},
\end{eqnarray}
 with $\kappa_{\bot}$ the perpendicular conductivity and $\kappa_{\|}$
 the parallel conductivity and $\hat{\bm{b}}$ the unit vector along
 the magnetic field \citep{braginskii65}. We will focus here on this
 equation, but one has to keep in mind that the diffusion of cosmic
 rays has a similar form
\begin{eqnarray}
\frac{\partial e_{cr}}{\partial t}=\bm{\nabla}.\left(D_{\|}\hat{\bm{b}}\hat{\bm{b}}.\bm{\nabla}e_{cr}\right)+\bm{\nabla}.\left[D_{\bot}({{\bm I}}-\hat{\bm{b}}\hat{\bm{b}})\bm{\nabla}e_{cr}\right],
\end{eqnarray}
with $D_{\bot}$ the perpendicular diffusion coefficient and $D_{\|}$
the parallel diffusion coefficient. This equation is therefore
solved by the same subroutine.

In cluster of galaxies the ion giroradius is much smaller than the mean free path between particle collisions and therefore the perpendicular part could be neglected since $\kappa_{\bot} \ll \kappa_{\|}$ (and $D_{\bot} \ll D_{\|}$ for cosmic rays). The conduction is highly anisotropic, primarily along the magnetic field, and mainly due to electrons. We adopt here the Spitzer value $\kappa_s$ for the conductivity \citep{spitzer62} of an ionised plasma with the Coulomb logarithm $\textrm{ln}\lambda$ set to a typical value for clusters:
\begin{eqnarray}
\kappa_{\|}=\kappa_S=9.2\times 10^{30}n_ek_B \left( \frac{k_B T}{\textrm{5~keV}} \right) ^{\frac{5}{2}}\left(\frac{n_e}{\textrm{0.01~cm$^{-3}$}}\right)^{-1}\left(\frac{37}{\textrm{ln}\lambda}\right)~\textrm{cm$^2$.s$^{-1}$}
\end{eqnarray}
For the parallel diffusion coefficient of the cosmic rays, we set
\begin{eqnarray}
D_{\|} = 10^{29}\textrm{~cm$^2$.s$^{-1}$}.
\end{eqnarray}


\subsection{Implementation and tests}
\subsubsection{Centered asymmetric method: advantages and drawbacks}

The first method we implemented uses the so-called centered asymmetric differencing. It is the most natural conservative discretization and it has been shown to give good results by \citet{parrish05,parrish07}. The idea is to compute the heat flux $F$ on each face and then to evolve the energy using an explicit time integration. We will consider only two dimensions but the generalisation to three dimensions is straightforward. The update of the energy is,
\begin{eqnarray}
-\frac{e^{n+1}_{i,j}-e^{n}_{i,j}}{\Delta t}=\frac{F^n_{i+1/2,j}-F^n_{i-1/2,j}}{\Delta x}+\frac{F^n_{i,j+1/2}-F^n_{i,j-1/2}}{\Delta y}.
\end{eqnarray}

\begin{figure}[h!]
  \plotone{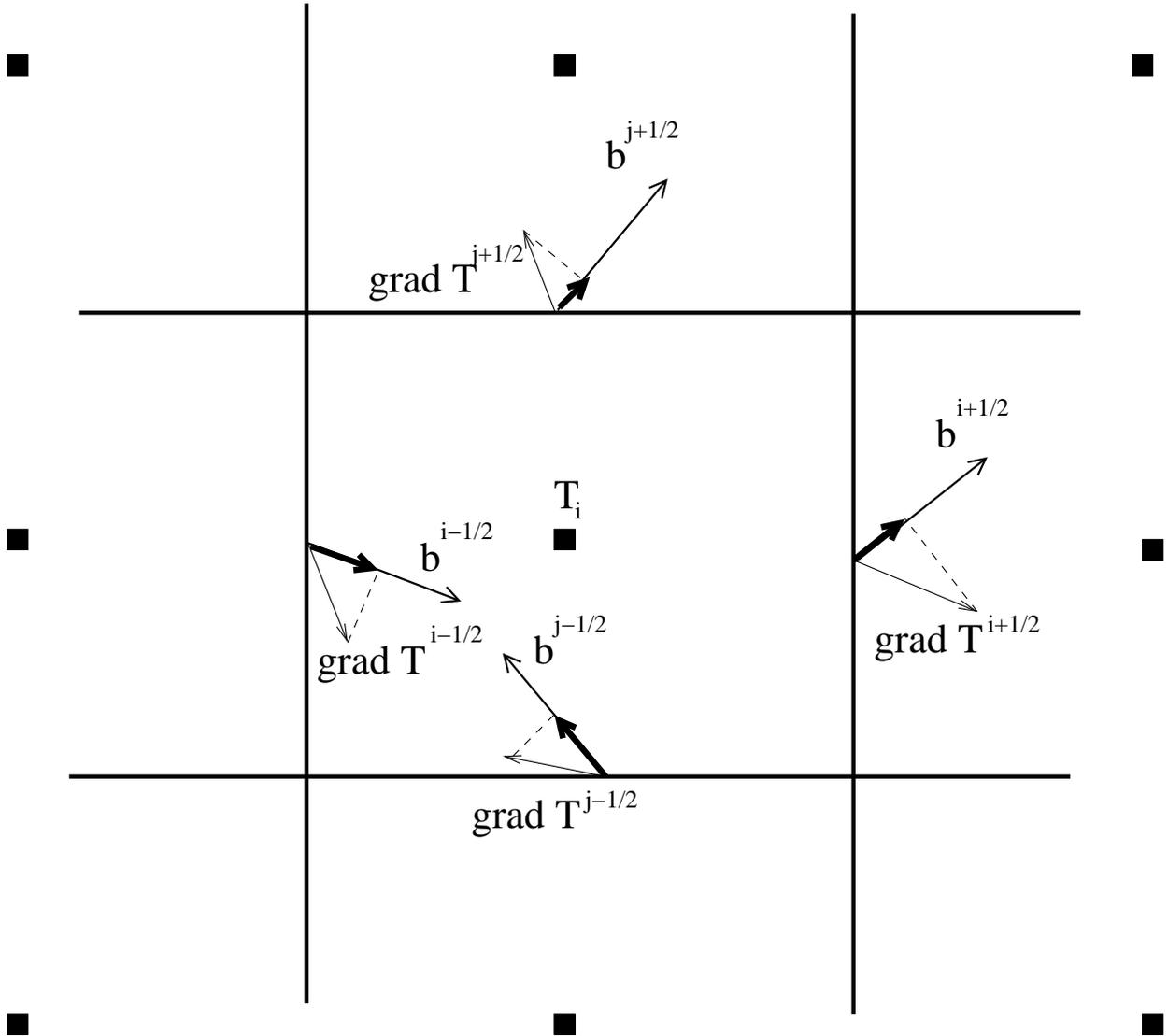}
    \caption{In the centered asymmetric method, the heat fluxes on each faces (bold arrows) are computed by projecting temperature gradients on the magnetic field. Temperature gradients computation require the knowledge of the neighbours (squares).}
    \label{gridasym}
  \end{figure}

This is the starting point for any conservative methods, now the problem is to evaluate the face-centered flux. The flux at time $n$ and position $(i+1/2,j)$ is given by (see Fig.\ref{gridasym}),
\begin{eqnarray}
F^n_{i+1/2,j}&=&b_x \bar{\kappa}_{\|} (b_x \frac{\partial T}{\partial x}+\bar{b}_y \frac{\bar{\partial T}}{\partial y}),\\
b_x&=&b^n_{x,i+1/2,j},\\
\bar{\kappa}_{\|}&=&\frac{\kappa_{i,j}+\kappa_{i+1,j}}{2},\\
\frac{\partial T}{\partial x}&=&\frac{T_{i+1,j}-T_{i,j}}{\Delta x},\\
\bar{b}_y&=&\frac{b_{y,i,j-1/2}+b_{y,i,j+1/2}+b_{y,i+1,j-1/2}+b_{y,i+1,j+1/2}}{4}\\
\frac{\bar{\partial T}}{\partial y}&=&\frac{T_{i+1,j+1}-T_{i+1,j-1}+T_{i,j+1}-T_{i,j-1}}{4\Delta y}.
\end{eqnarray}
The x component of the temperature gradient and the magnetic field are well known in $(i+1/2,j)$, however the y components need to be extrapolated (overline). The time step is choosen to ensure linear stability
\begin{eqnarray}
\Delta t=0.45 \textrm{~min~}\left( \frac{\Delta x^2}{2 D_{\textrm{cond}}} \right),\\
D_{\textrm{cond}}=(\gamma-1) \frac{\kappa_{\|}}{\rho}\frac{\mu m_H}{k_B}.
\end{eqnarray}
This method is fast, efficient and accurate. However, as highlighted by \citet{sharma07}, this method is \emph{not positive definite}. Indeed, it could lead to negative temperature in presence of large temperature gradient. An easy way to see the problem is to notice that in the flux expression $T_{i+1,j+1}$ and $T_{i,j+1}$ appear with positive signs. So if one of this temperature is a lot larger than all the others then nothing could balance this very large negative contribution and the energy $e^{n+1}_{i,j}$ could become negative. This problem is due to the spatial discretization itself and not to the explicit scheme used for time integration. The transverse temperature gradient is not computed from the same origin as where the energy is taken, this is the heart of the problem. An implicit scheme \citep{sovinec05, balsara08} could therefore also suffer from the same negative temperature issues. One could indeed imagine the same situation as before but where the very large temperature $T_{i+1,j-1}$ or $T_{i,j-1}$ stays relatively constant until the time step $t^{n+1}$. Then, the energy $e^{n+1}_{i,j}$ could also become negative with an implicit scheme. 

There are two methods to make the scheme positive: the first one consists in limiting the transverse gradient $\frac{\partial T}{\partial y}$. This idea is described in full detail in \citet{sharma07}. The second one consists in another discretization of the problem and is described in the following section.

\subsubsection{Positive anisotropic heat conduction: the flux-tube method}

We present here a new positive method for anisotropic conduction. It is based on a physically motivated discretization which treats anisotropic conduction as a 1D diffusion process along the field lines. One could indeed simplify the discretization by considering only one thin magnetic flux-tube containing $(i,j)$ and by calculating the temperature gradient and energy flux directly along this flux tube. Using $\bm{\nabla}.\bm{B}=0$, the anisotropic conduction equation
\begin{eqnarray}
  \frac{\partial e}{\partial t}&=&\bm{\nabla}\left(\frac{\bm{B}}{B}\kappa_{\|}\hat{\bm{b}}.\bm{\nabla}T\right)
\end{eqnarray}
could be rewritten as
\begin{eqnarray}
 \frac{\partial e}{\partial t}&=&\bm{B}.\bm{\nabla}\left(\frac{\kappa_{\|}}{B}~\hat{\bm{b}}.\bm{\nabla}T\right),
\end{eqnarray}
or,
\begin{eqnarray}
  \frac{1}{B}\frac{\partial e}{\partial t}&=&\hat{\bm{b}}.\bm{\nabla}\left(\frac{\kappa_{\|}}{B}~\hat{\bm{b}}.\bm{\nabla}T\right),
\end{eqnarray}
where $\hat{\bm{b}}=\bm{B}/B$. One can see the apparition of the derivative along the magnetic field, and the term $1/B$ to satisfy magnetic flux conservation. Because $B.A=\textrm{constant}$, where $A$ is the cross-section area of a flux-tube, the $1/B$ term can be thought of as representing the cross-sectional area $A$.

We now define $s$ as the curvilinear abscissa along the field lines. The origin of this curvilinear abscissa is choosen to be $s=0$ at the grid point $(i,j)$ of interest. In order to compute derivatives, we consider variations over a length $\Delta s=\Delta x$ along the field line. The derivative of a function $f_s$ in $(i,j)$ becomes $(f_{+\Delta s /2}-f_{-\Delta s /2})/\Delta s$ and the same function derived in $+\Delta s /2$ gives $(f_{+\Delta s}-f_{0})/\Delta s$.

\begin{figure}[h!]
  \plotone{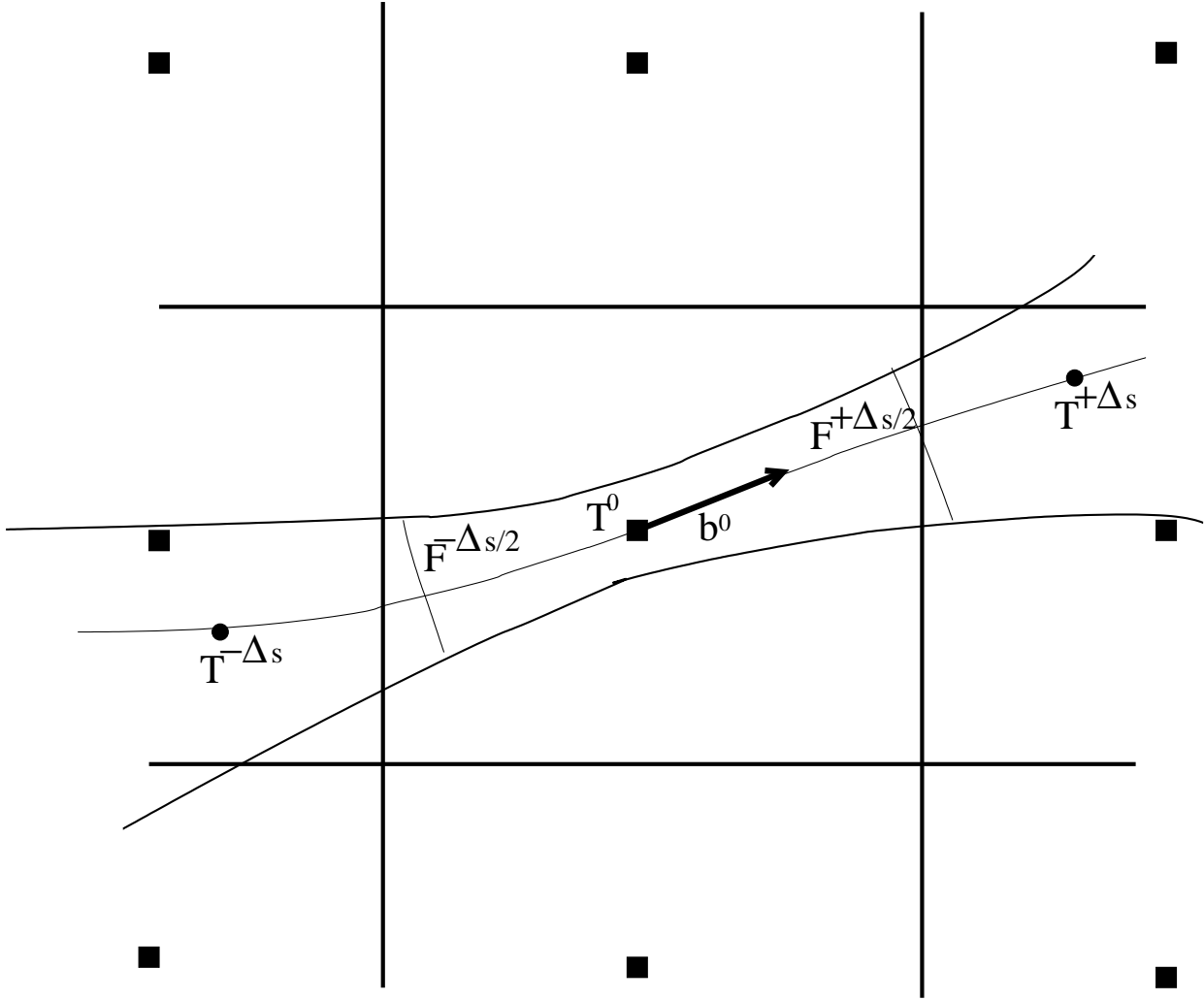}
  \caption{In the flux-tube method, one considers a thin magnetic flux tube around $T_0$. The evolution of this temperature depends on the heat flux along the magnetic field which is evaluated on the two faces at a distance $\Delta s/2$ from the center: $F^{\pm \Delta s /2}$. These fluxes are deduced from the temperatures $T^{\pm\Delta s}$ which are themselves computed using all the neigbhours (squares).}
    \label{gridfluxtube}
  \end{figure}

The discretization along the flux-tube is therefore (see Fig.\ref{gridfluxtube}),
\begin{eqnarray}
\frac{e^{n+1}_{0}-e^{n}_{0}}{\Delta t}&=&\frac{B_0}{\Delta s}\left[ \left(\frac{\kappa_{\|}}{B}\frac{\partial T}{\partial s}\right)_{+\Delta s/2}-\left(\frac{\kappa_{\|}}{B}\frac{\partial T}{\partial s}\right)_{-\Delta s/2}\right],\\
\left(\frac{\partial T}{\partial s}\right)_{+\Delta s/2}&=&\frac{\bar{T}_{+\Delta s}-T_0}{\Delta s},\\
\left(\frac{\partial T}{\partial s}\right)_{-\Delta s/2}&=&\frac{T_0-\bar{T}_{-\Delta s}}{\Delta s},\\
\kappa_{\|,-\Delta s/2}&=&\frac{\bar{\kappa}_{\|,-\Delta s}+\kappa_{\|,0}}{2},\\
\kappa_{\|,+\Delta s/2}&=&\frac{\bar{\kappa}_{\|,+\Delta s}+\kappa_{\|,0}}{2},\\
B_{-\Delta s/2}&=&\frac{\bar{B}_{-\Delta s}+B_{0}}{2}\label{averageBm},\\
B_{+\Delta s/2}&=&\frac{\bar{B}_{+\Delta s}+B_{0}}{2}\label{averageBp},
\end{eqnarray}
with the subscript indicating the curvilinear abscissa where the value is computed and the overline meaning that the value is not directly known and has therefore to be interpolated from grid point values.

The next step is to estimate the position $\bm{X}(s)$ corresponding to $s=\pm \Delta s$ in the cartesian grid, which is done at second order,
\begin{eqnarray}
\bm{X}(s)&=&\hat{\bm{b}}_0 s+0.5 s^2 (\hat{\bm{b}}_0.\nabla)\hat{\bm{b}}_0,\\
(\hat{\bm{b}}_0.\nabla)\hat{\bm{b}}_{x,0}&=&b_{x,i,j}\frac{b_{x,i+1,j}-b_{x,i-1,j}}{2 \Delta x}+b_{y,i,j}\frac{b_{x,i,j+1}-b_{x,i,j-1}}{2 \Delta x},\\
(\hat{\bm{b}}_0.\nabla)\hat{\bm{b}}_{y,0}&=&b_{x,i,j}\frac{b_{y,i+1,j}-b_{y,i-1,j}}{2 \Delta x}+b_{y,i,j}\frac{b_{y,i,j+1}-b_{y,i,j-1}}{2 \Delta x}.
\end{eqnarray}

The final and fundamental step is the interpolation of the temperature at the curvilinear abscissa $s=\pm \Delta s$. This will determine the accuracy of the method, as well as the positivity of the scheme. For this purpose, we decompose the 2D interpolation into a series of 1D interpolations that are done with the second-order Lagrange interpolating formula. An important point, is that whatever the position we consider, we only interpolate using $(i,j)$ and the 8 surrounding points. This is more convenient for the boundary conditions. Unfortunately, second-order interpolations are not guaranteed to stay in the range defined by the 2 extrema of the 9 considered point. Allowing such overshoot could create oscillations and negative temperature. We therefore saturate the interpolation to the extrema of the 9 considered point, in order to allow positivity of the scheme. One drawback is that we loose accuracy near extrema, but this is unavoidable in order to get physical results. We will also see that the resulting amount of artificial diffusion is small. Finally, the norm of the magnetic field and the conductivity are interpolated only at first order for speed and therefore don't need to be saturated. One could interpolate at higher order for better accuracy.

It is interesting to note that one could rewrite the update of the energy as,
\begin{eqnarray}
T^{n+1}_{0}&=&T^n_0+\alpha \left( \frac{A_{-\Delta s/2}T_{-\Delta s}+A_{+\Delta s/2}T_{+\Delta s}}{A_{-\Delta s/2}+A_{+\Delta s/2}}-T^n_0\right),\\
A_{-\Delta s/2}&=&\frac{\kappa_{\|,-\Delta s/2}}{\kappa_{\|,0}} \frac{B_0}{B_{-\Delta s/2}},\\
A_{+\Delta s/2}&=&\frac{\kappa_{\|,+\Delta s/2}}{\kappa_{\|,0}} \frac{B_0}{B_{+\Delta s/2}},\\
\alpha&=&D_{\textrm{cond},0} \frac{\Delta t}{\Delta s^2}(A_{-\Delta s/2}+A_{+\Delta s/2}),\\
D_{\textrm{cond},0}&=&\kappa_{\|,0}\frac{\gamma-1}{\rho_0} \frac{\mu m_H}{k_B}.
\end{eqnarray}

It means that the temperature $T_0$ evolve by a fraction $\alpha$ toward the arithmetic average of $T_{-\Delta s}$ and $T_{+\Delta s}$. We therefore choose the time step like in the precedent method that is to say, $\Delta t=0.45 \textrm{~min~}(\Delta x^2/(2 D_{\textrm{cond}}))$. Using this time step and computing the norm of the magnetic field by \ref{averageBm} and \ref{averageBp}, it guarantees that $\alpha \le 1$ and prevents from any overshoot of the average. Since the interpolated temperature $T_{-\Delta s}$ and $T_{+\Delta s}$ are between the extrema of the neighbors of $T_0$, it means that no oscillations could appear! We have therefore implemented a positive flux-tube scheme for anisotropic conduction and diffusion.

\subsubsection{Diffusion of a step function}

The first test we run is the passive diffusion of a 1D Heavyside function. The goal here is to check if the code solves well the diffusion along straight magnetic field lines. In a second test we will check how well the code follows curved magnetic field line. We start here with the following conditions,
\begin{eqnarray}
\rho=1, b=1 &\textrm{everywhere}&\\
e=1&\textrm{~for~}& x\le 0.5\\
e=2&\textrm{~for~}& 0.5 < x\le 0.75\\
e=1&\textrm{~for~}& x> 0.5
\end{eqnarray}

\begin{figure}[h!]
  \plotone{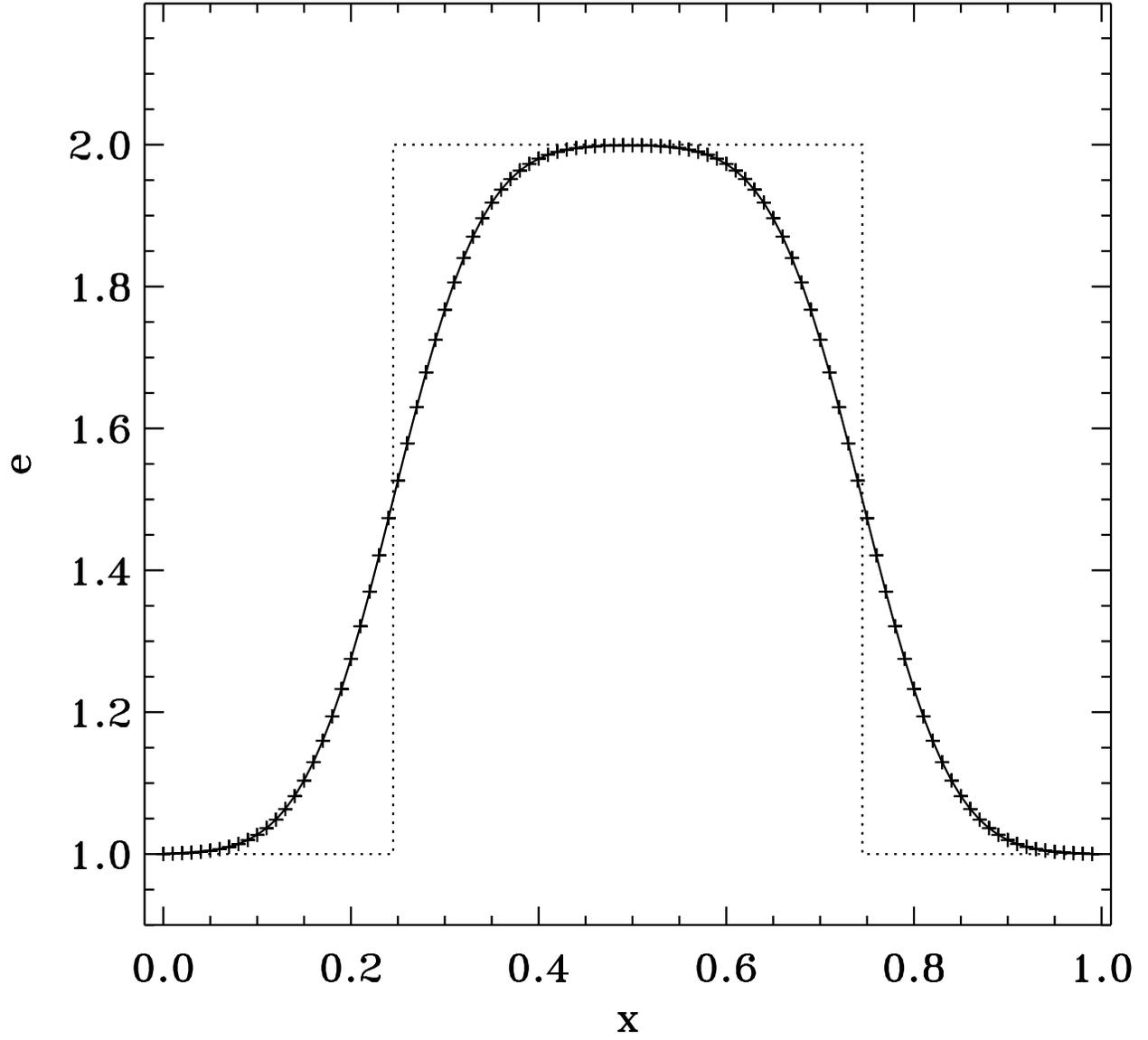}
    \caption{Diffusion of 2 1D Heavyside functions (dotted line). The simulation points (plus signs) match well with the analytical result (line).}
    \label{heavy}
  \end{figure}

We then use a constant conduction coefficient, $D_{\textrm{cond}}=1$ so that the solution is analytically tractable. Our 100 grid points simulation is ran until $t=2.8\times 10^{-3}$. For one step of size $\Delta e$ and mean $e_0$ situated at the location $x_0$ the analytical solution gives,
\begin{eqnarray}
e(x,t)=e_0+\frac{\Delta e}{2} \textrm{erf}\left(\frac{x-x_0}{\sqrt{4D_{\textrm{cond}}t}}\right).
\end{eqnarray}
The comparison in Fig.\ref{heavy} shows a very good agreement between the simulation and the analytical solution since we cannot differenciate them.

\subsubsection{Anisotropic conduction in circular magnetic field lines}

A more challenging test involves passive anisotropic diffusion along circular field lines, as proposed in \citet{parrish05}. The idea is to consider an initial hot patch embedded in circular magnetic field lines. The heat should then diffuse along the field lines but not across the field. We start here with the following initial condition:
\begin{eqnarray}  
\rho=1 &\textrm{for}& 0\le x\le 1 \textrm{~and~} 0 \le y \le 1,\\
b_x=\frac{y-0.5}{r} &\textrm{for}& 0 \le x\le 1 \textrm{~and~} 0 \le y \le 1,\\
b_y=-\frac{x-0.5}{r}&\textrm{for}& 0 \le x \le 1 \textrm{~and~} 0 \le y \le1,\\
e=10000         &\textrm{for}& 0.7\le x \le 0.8 \textrm{~and~} 0.49\le y \le 0.51,\\
e=1             &\textrm{otherwise},&
\end{eqnarray}
with $r=\sqrt{(x-0.5)^2+(y-0.5)^2}$.

\begin{figure*}                                  \begin{tabular}{ccc}
   & \includegraphics[width=0.3\hsize]{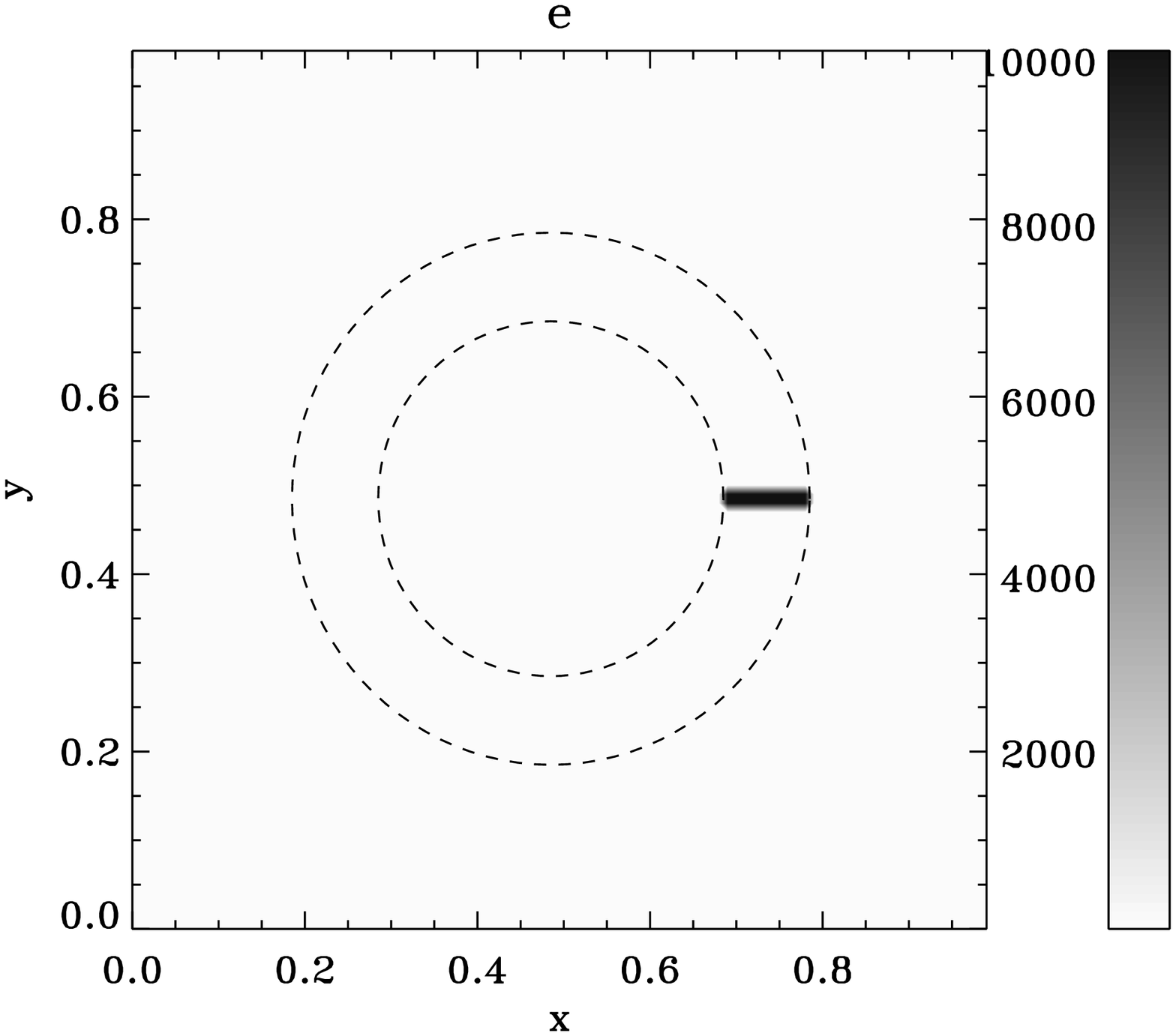}&\\
\includegraphics[width=0.3\hsize]{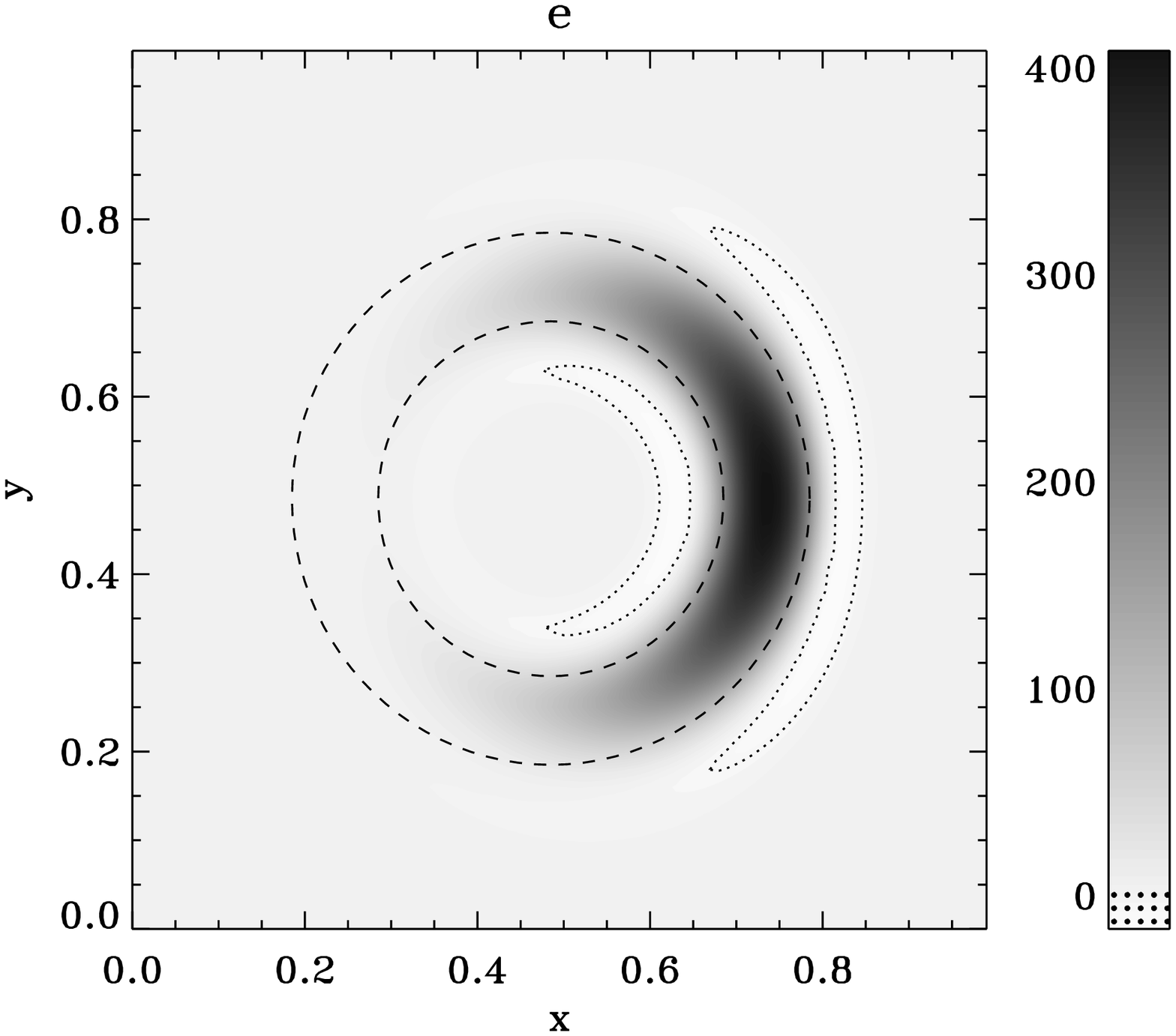}&\includegraphics[width=0.3\hsize]{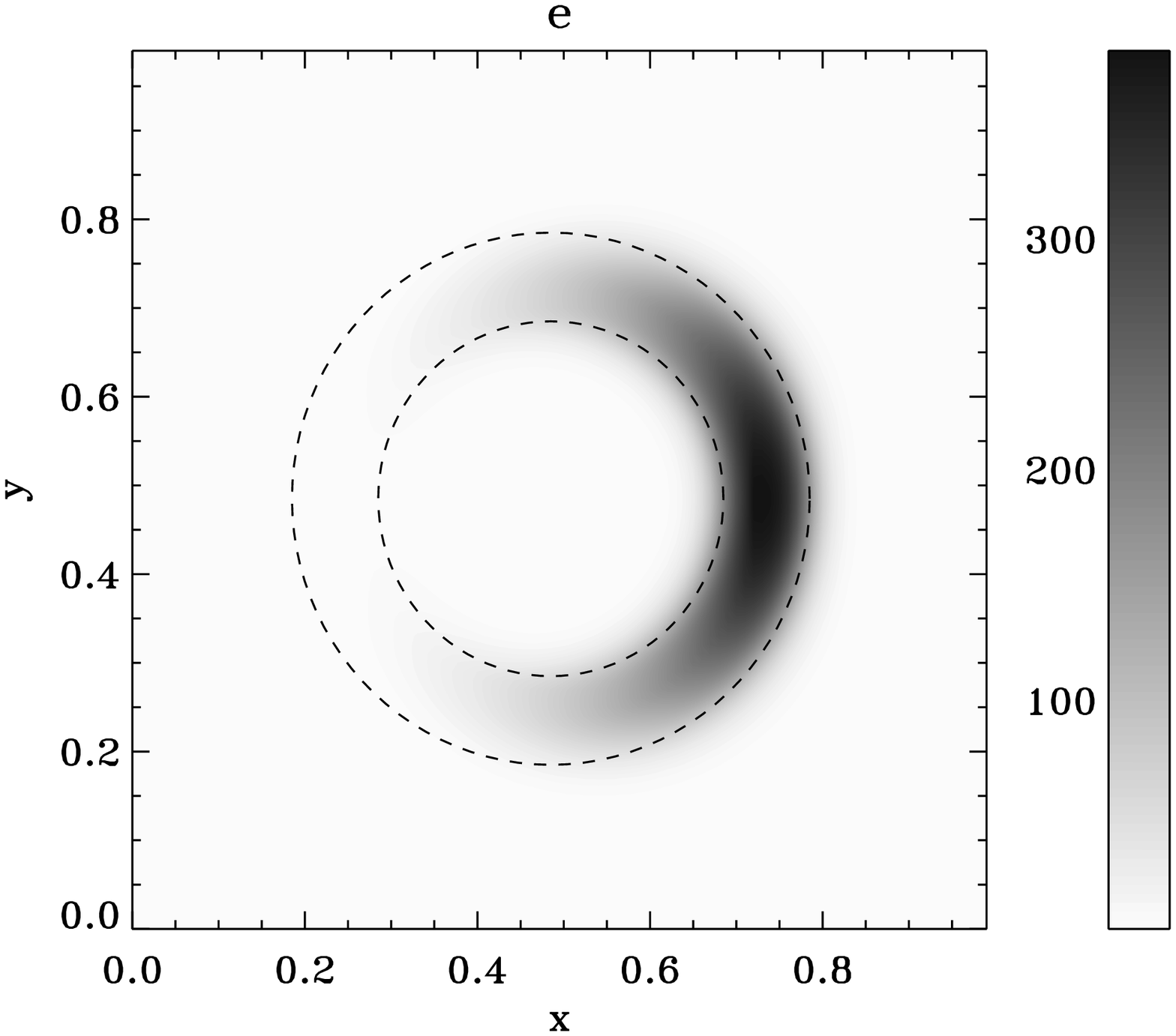}&\includegraphics[width=0.3\hsize]{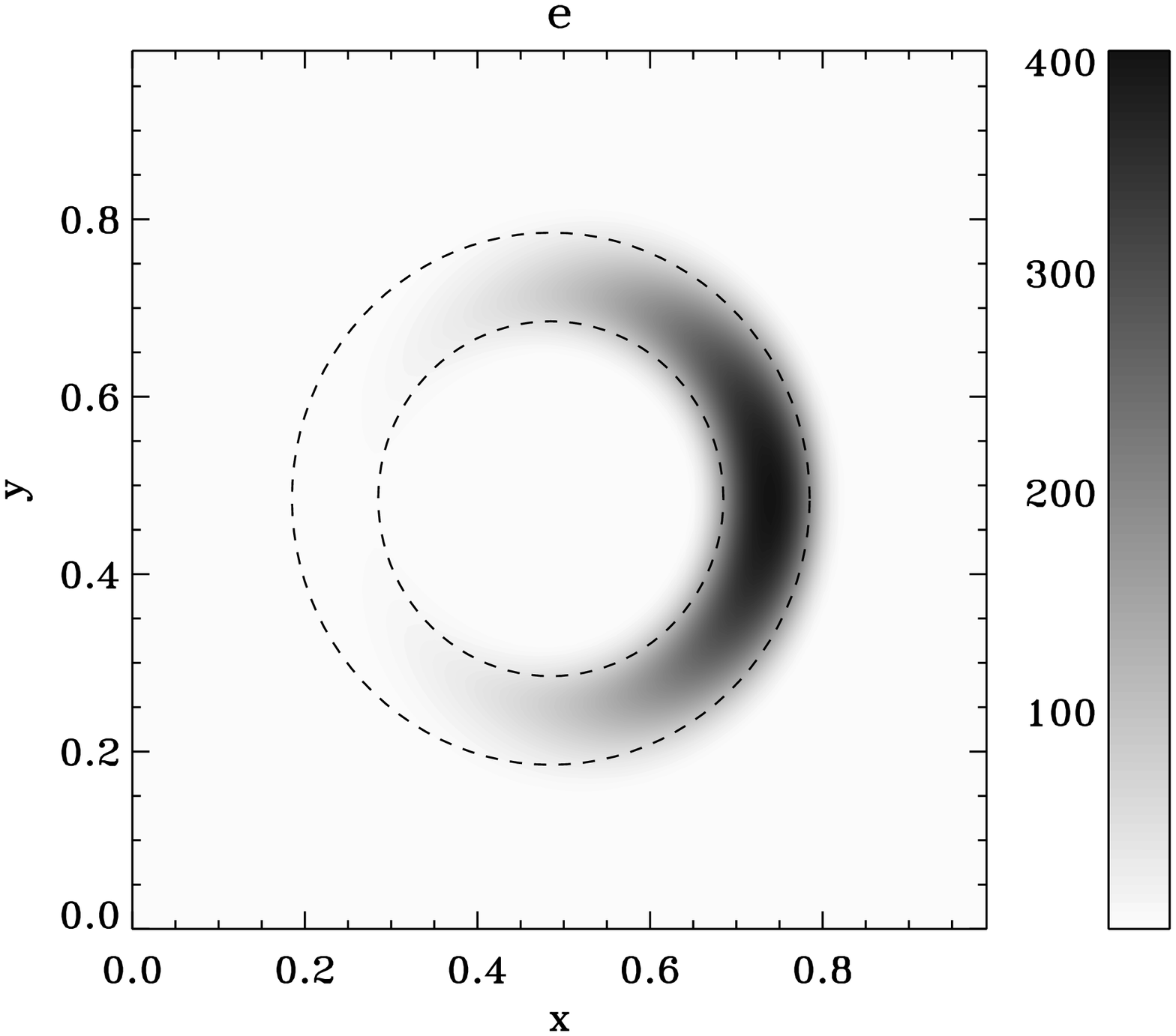}\\
\includegraphics[width=0.3\hsize]{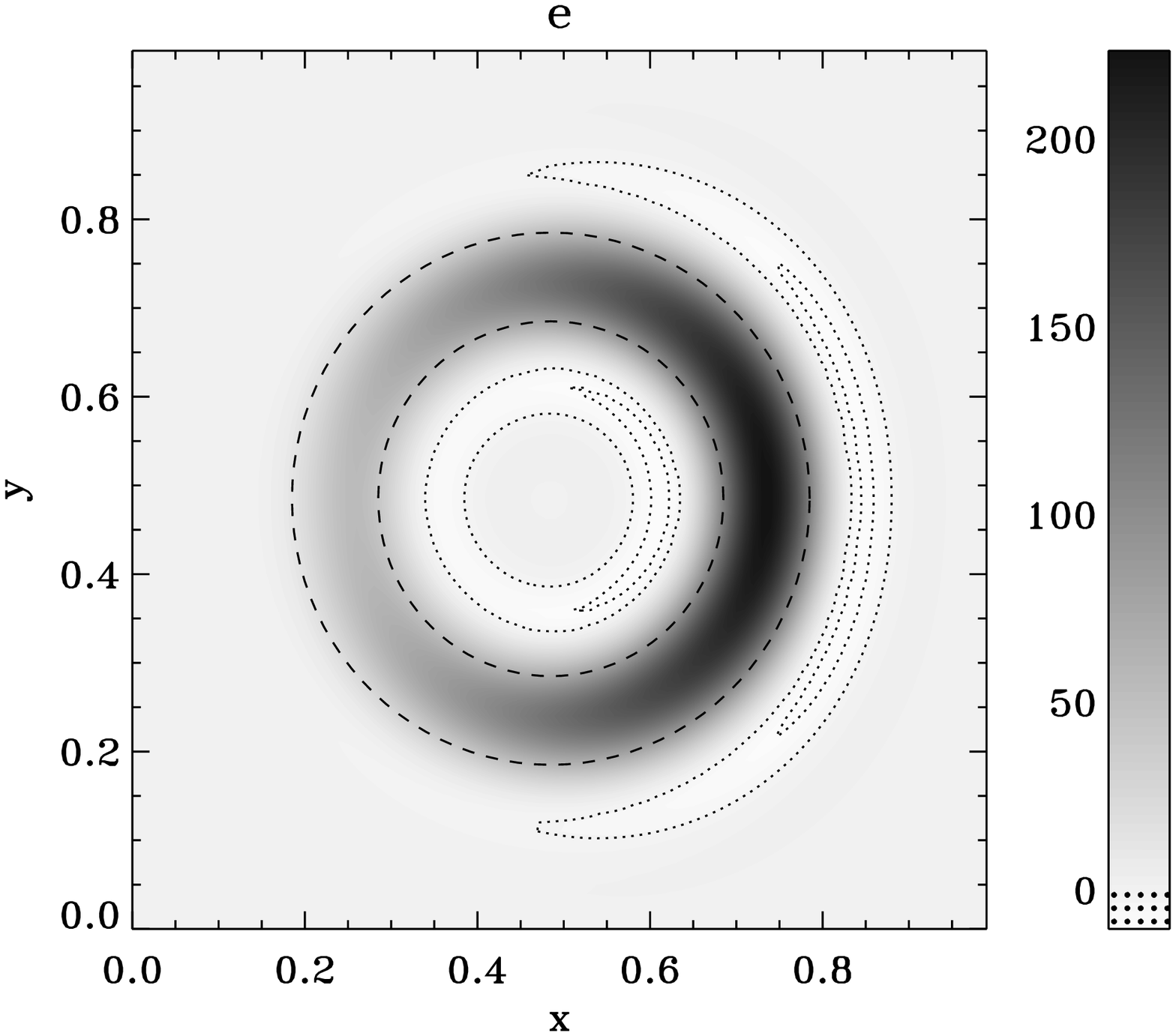}&\includegraphics[width=0.3\hsize]{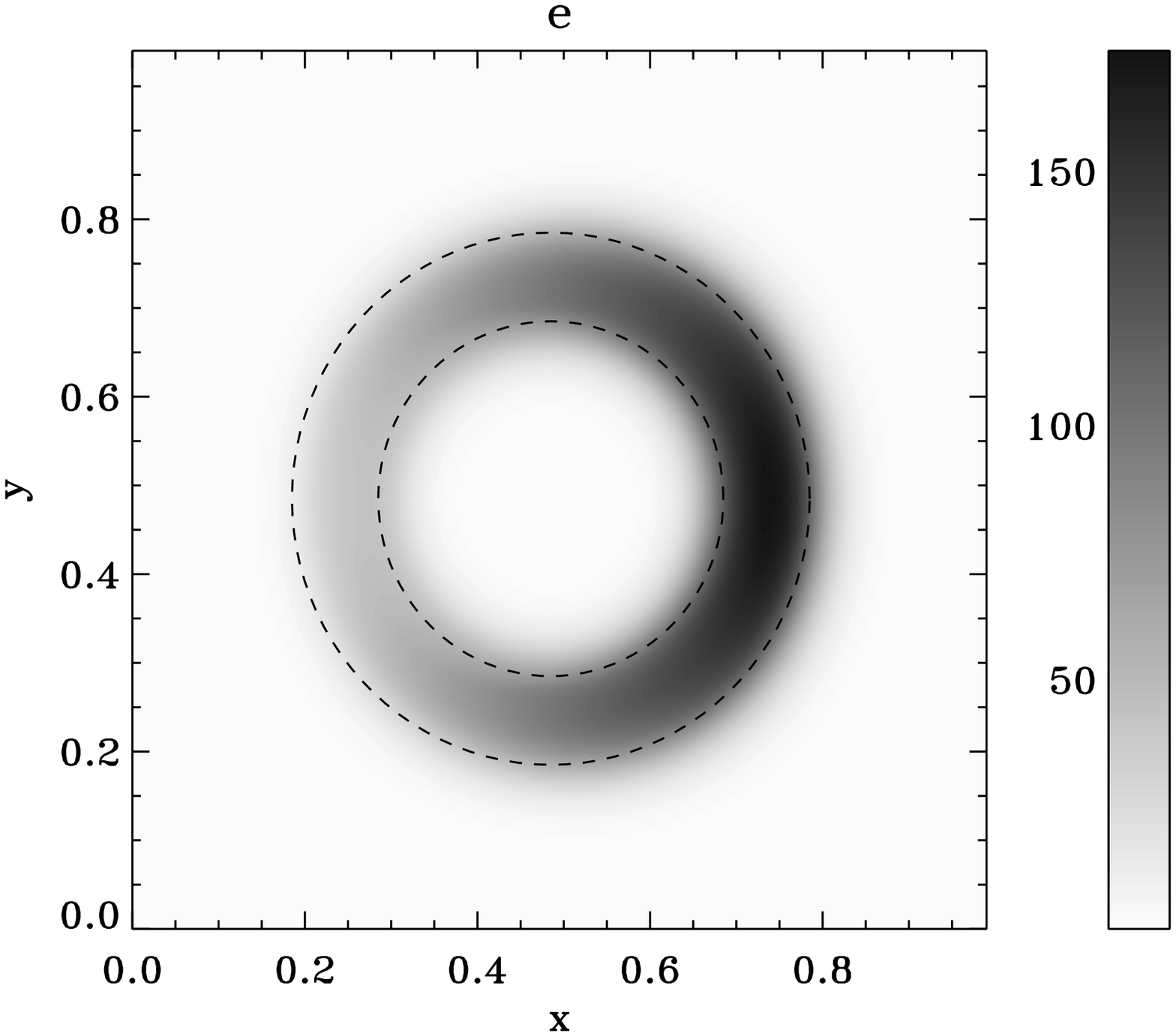}&\includegraphics[width=0.3\hsize]{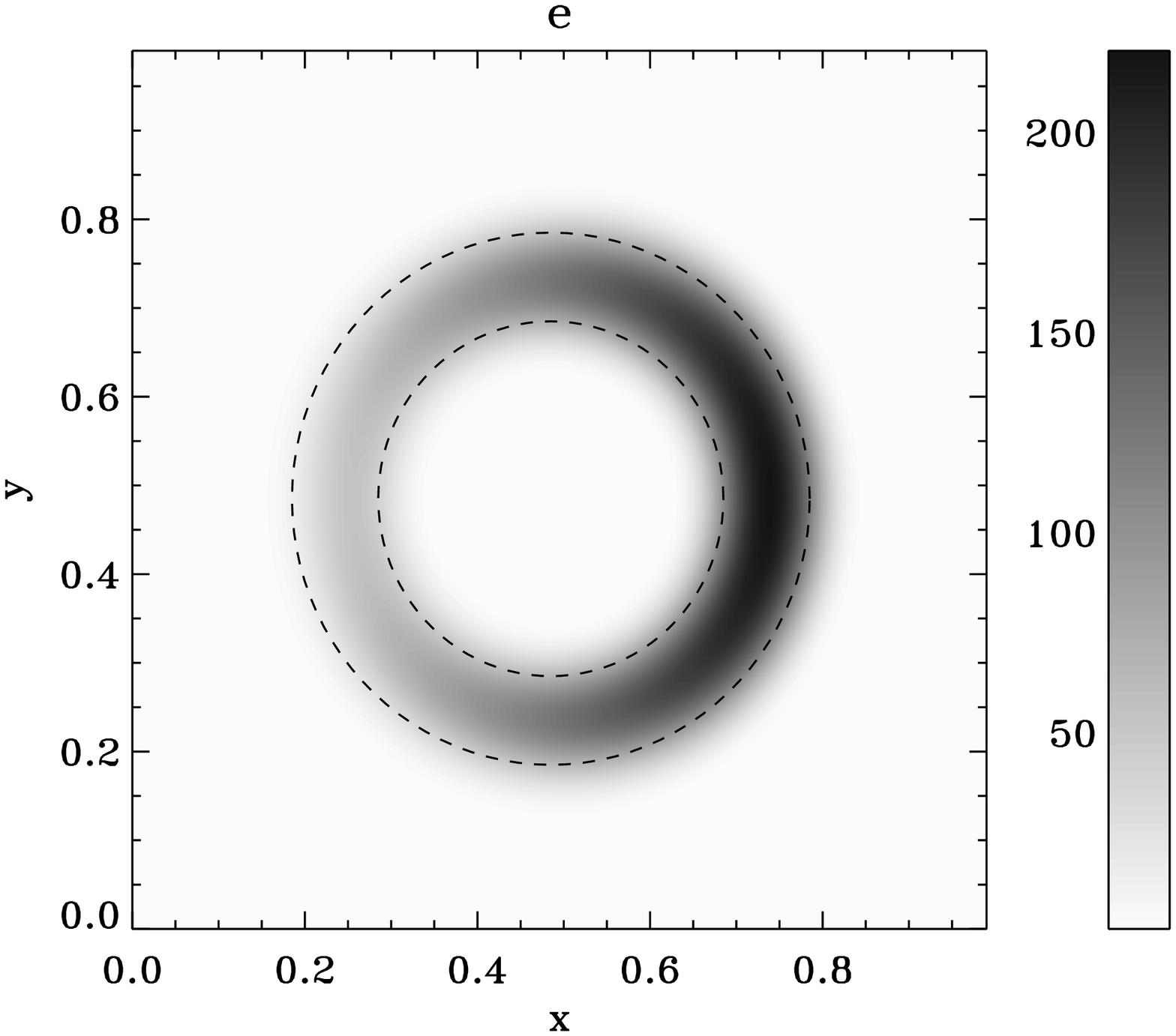}\\
\includegraphics[width=0.3\hsize]{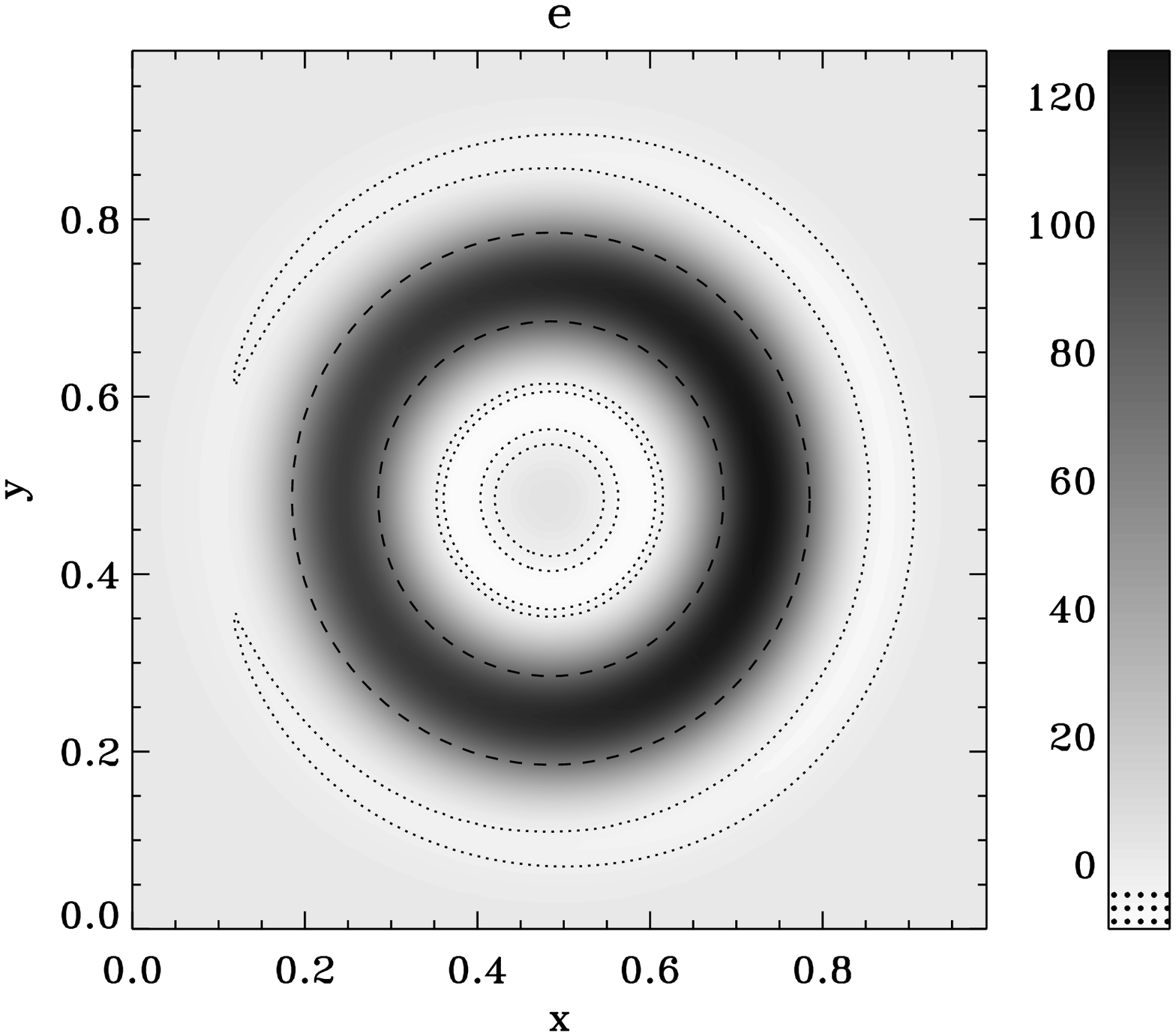}&\includegraphics[width=0.3\hsize]{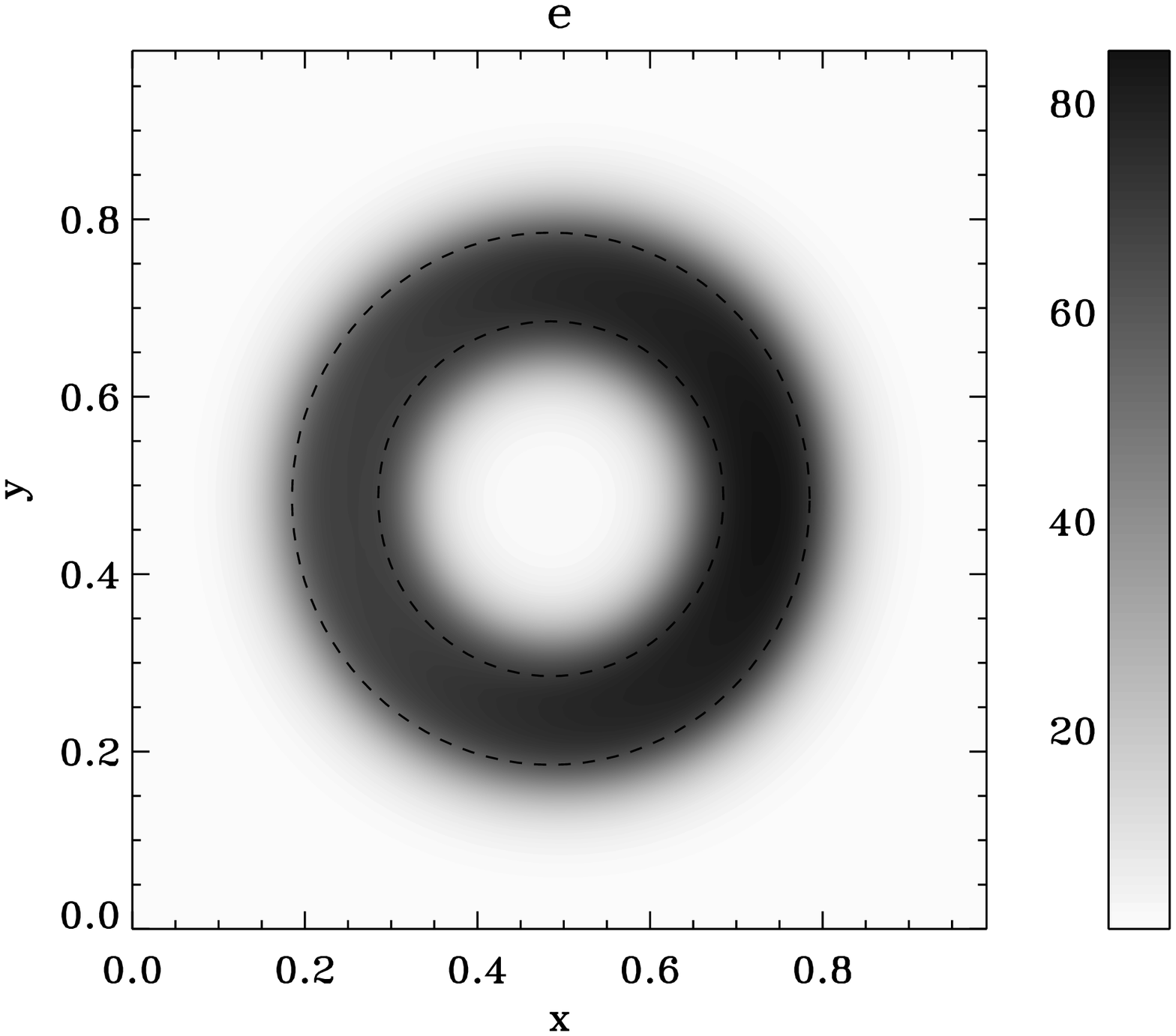}&\includegraphics[width=0.3\hsize]{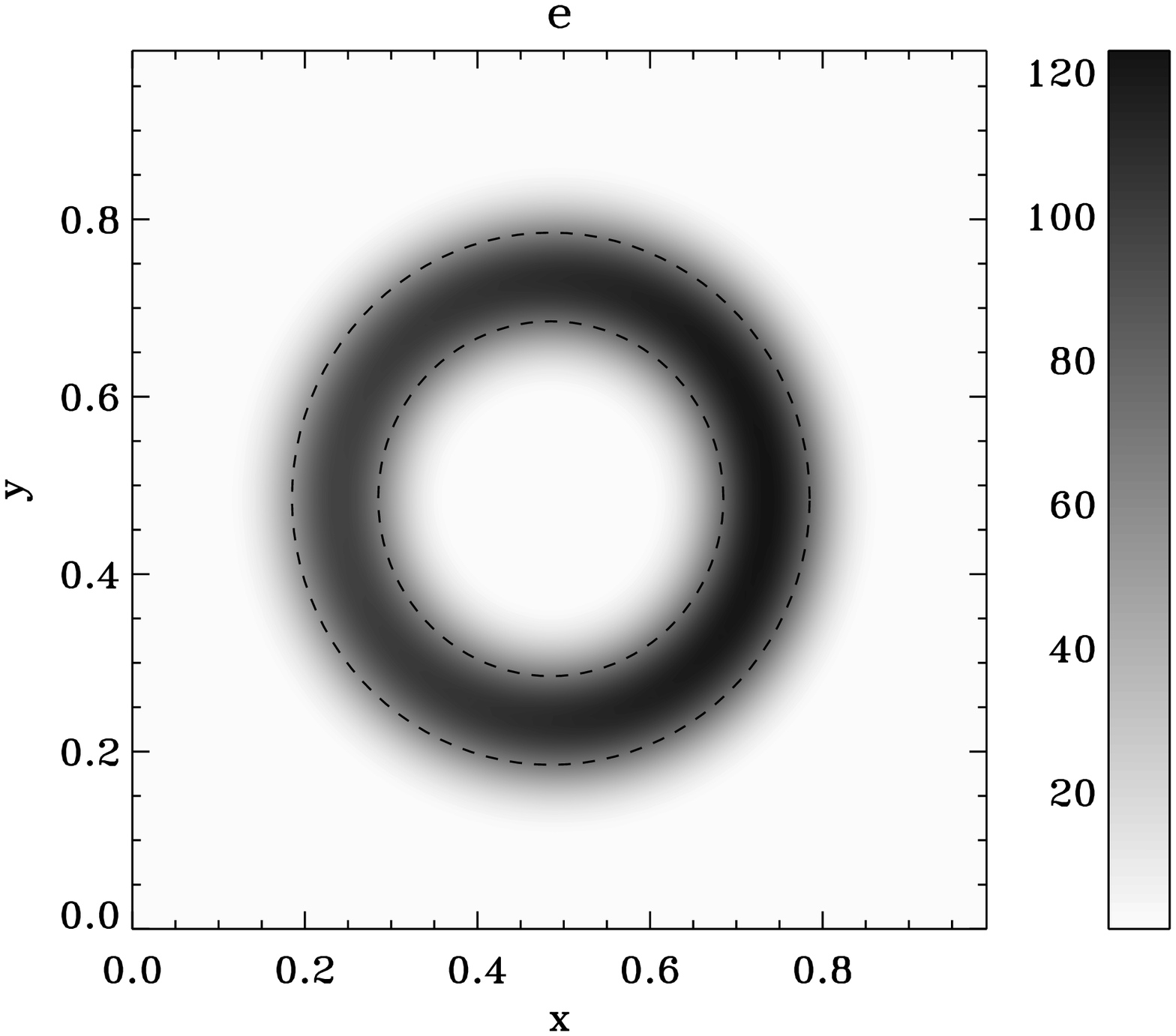}\\
  \end{tabular}  
  \caption{Anisotropic diffusion of a hot patch (first graph from the top) along circular magnetic field in a simulation with $100^2$ grid points. Columns correspond respectively to, the standard centered asymmetric discretization, the Van-Leer limited methods and our new flux-tube method. Lines correspond respectively to $t=0.0225$, $t=0.0675$ and $t=0.18$. The circles are the field lines confining the heat. Note the negative temperature (dotted contours) in the first method and the considerable perpendicular diffusion in the second method. \label{circle}}
\end{figure*}

The $100$ by $100$ simulation is run with $D_{\textrm{cond}}=1$ using our flux-tube method as well as the standard centered asymmetric method. We also run the Van-Leer-limited implementation of \citet{sharma07}. We present in Fig.\ref{circle} the temperature profiles at $t=0$, $t=0.0225$, $t=0.0675$ and $t=0.18$ for these three methods.

In all three methods, the heat flux follows the circular field lines and tends toward a stationnary solution without any angular gradient of temperature. When there is no perpendicular conductivity, the analytical stationnary solution is obtained by energy conservation: $e_{stat}=128.3$ everywhere inside the shell. However, second order truncation errors add some artificial perpendicular diffusion. As a consequence the radial profile which was initially a Heavyside (as in the preceding test) diffuses. The consequence is that the maximum is lowered and the radial profile is smoothed.

If one uses the standard asymmetric differencing, the dramatic consequence is that it leads to negative temperatures, even after a long run time. These negative temperatures are shown as a white inner and outer circle with dotted contours in Fig.\ref{circle}, left column. This is a very important problem. On the numerical point of view, it indicates that this scheme could overshoot the extrema and therefore create some spurious oscillations. On the physical point of view, it means that heat can flow from lower to higher temperature. Moreover, while coupling with the MHD solver, it could lead to negative temperatures, create an imaginary sound speed, and lead to unphysical results.

All these points have been discussed in detail in \citet{sharma07}. They found that this problem arises in presence of strong temperature gradient perpendicular to the magnetic field. This is why they proposed a limited version of this asymmetric discretisation, in which they limit the perpendicular temperature gradient. However, as they already mentionned, the perpendicular diffusion becomes important if one uses limited methods. This is obvious, in Fig.\ref{circle}, middle column. For example, in the last line, one could see that the radial dispersion is larger than in the asymmetric method. Moreover, the maximum has been decreased by a factor of 1.5.

On the contrary, our method combines two advantages of the two other methods. As presented in the right-hand column of Fig.\ref{circle}, the temperature always stays between the initial extrema but keep a low level of perpendicular diffusion. We are now going to estimate this perpendicular numerical diffusion using a test especially dedicated for this purpose.

\subsubsection{Accuracy of the method: Sovinec test}

In order to compare the accuracy of different methods, it is interesting to know what is the artificial perpendicular diffusivity of a scheme. Indeed, some applications could require a large ratio of the parallel to perpendicular conductivity. \citet{sovinec05} have developped such a test. We will therefore run this test for our method and compare our perpendicular artificial diffusion with the Van-Leer-limited method presented in \citet{sharma07} as well as the standard asymmetric method.

The idea is to consider the full heat equation \ref{heat} in 2D with both a perpendicular and an anisotropic part. We also add in this energy equation a heating source term $Q(x,y)=Q_0 \times cos(k x) cos(k y)$. The equation to solve becomes,
\begin{eqnarray}
\frac{\partial e}{\partial t}=\bm{\nabla}.\kappa_{\|}\hat{\bm{b}}\hat{\bm{b}}.\bm{\nabla}T+\bm{\nabla}.\left[\kappa_{\bot}({{\bm I}}-\hat{\bm{b}}\hat{\bm{b}})\bm{\nabla}T\right]+Q(x,y).
\end{eqnarray}
The analytical stationnary solution could be computed in the case of pure isotropic conduction ($\kappa_{\|}=\kappa_{\bot}$) with a constant conductivity. It is given by 
\begin{eqnarray}
T(x,y)=\frac{Q_0}{2\kappa_{\bot}k^2} cos(k x) cos(k y). 
\end{eqnarray}

As in \citet{sovinec05}, we consider a fixed magnetic field satisfying $\bm{B}.\bm{\nabla} T=0$, so that the previous solution still apply. Taking into account artificial diffusion, the solution in the center becomes $T(0,0)=Q_0/\left[2k^2(\kappa_{\bot}+\kappa_{\textrm{num}})\right]$. The artificial diffusion could therefore easily be deduced from the central temperature.

Our initial conditions for $0.5 \le x \le 0.5$ and $0.5 \le y \le 0.5$ are,
\begin{eqnarray}
\rho&=&1\\
k&=&\pi\\
Q_0&=&2\pi^2\\
\kappa_{\bot}&=&0,\\
T(x,y)&=&cos(\pi x) cos(\pi y),\\
B_x&=&cos(\pi x) sin(\pi y),\\
B_y&=&-cos(\pi y) sin(\pi x).
\end{eqnarray}
We also choose $T_{bound}=0$ for the fixed boundary conditions.

\begin{figure}[h!]
  \plotone{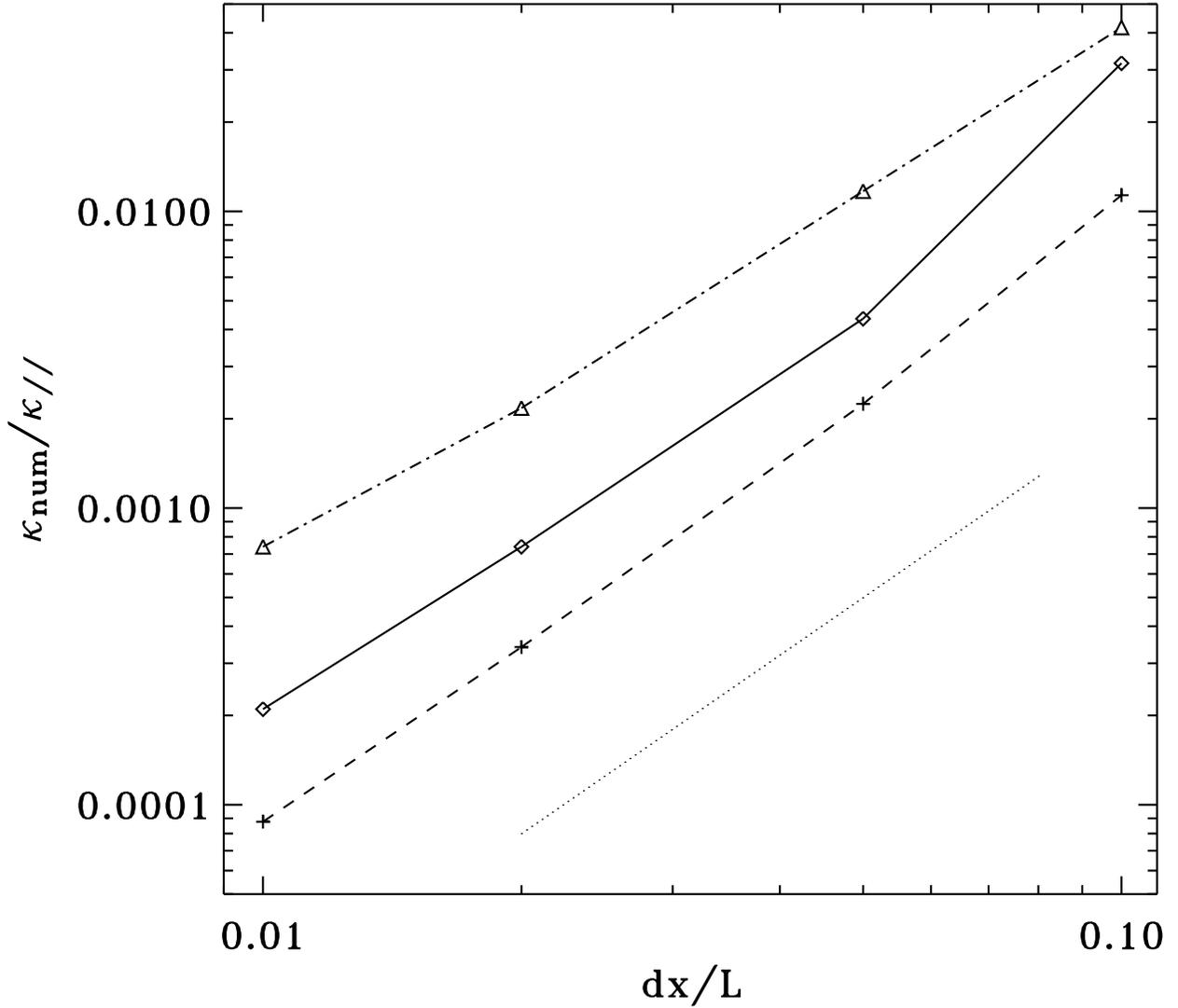}
    \caption{Measure of the perpendicular numerical diffusion $\kappa_{\textrm{num}}/\kappa_{\|}$ in the \citet{sovinec05} test as a function of the resolution. The dashed curve is the result for the standard asymmetric discretization. The dash-dotted curve is for the Van-Leer-limited method from \citet{sharma07}. Finally, the continuous curve is for our new flux-tube method. For reference, the dotted line has a slope of 2.}
    \label{numdiff}
  \end{figure}

Unlike \citet{sharma07} and \citet{sovinec05}, our goal here, is to estimate the numerical diffusion in the case of a pure anisotropic conduction ($\kappa_{\bot}=0$). This numerical diffusion increases with the parallel conductivity with $\kappa_{\bot}=0$. We obtain $\kappa_{num}$ by running simulation to steady state and setting $\kappa_{num}=1/T(0,0)$. Using this method, we determine the ratio $\kappa_{num}/\kappa_{\|}$ as a function of the resolution ($dx/L=0.01$, $dx/L=0.02$, $dx/L=0.05$ and $dx/L=0.1$) for the different implementations of the conduction. The results are presented on Fig.\ref{numdiff}.

The first point is that all the methods converge towards lower numerical diffusion with an order of convergence of $\approx 2$ (i.e., $\kappa_{num}/\kappa_{\|}\propto dx^2$). The least diffusive method is of course the standard asymmetric method. This method reaches a ratio of $\kappa_{num}/\kappa_{\|}=10^{-4}$ for a resolution $dx/L=0.01$. However, we have already noted in the precedent part that this method could lead to unphysical results and may therefore not be suitable for applications with large temperature gradients (like in presence of shocks). The method of \citet{sharma07} circumvents this problem but the perpendicular diffusion is a factor of $\approx 7$ more important than in the standard method. On the contrary our new method is only a factor of $\approx 2$ more diffusive than the standard method but doesn't lead to negative temperature.

\subsection{Conclusion: comparison of the three methods for anisotropic conduction or diffusion}

We summarize in Table~\ref{methods_anis} the properties of the three different methods studied in this section. One of the most important property emphasized by \cite{sharma07} is to know if the solution remains bounded in the initial range of temperature. Indeed, this is essential to guarantee physical results and stability of the scheme. Unfortunately, the standard asymmetric method doesn't share this property. It still could be used in presence of smooth temperature field taking advantage of its speed (4 times faster than the MHD solver) and accuracy ($\kappa_{num}/\kappa_{\|}= 10^{-4}$) but has to be avoided in presence of strong temperature gradient and chaotic magnetic field.

From our knowledge, only two methods for asymmetric conduction (or
diffusion) are positive definite. In the first method, an asymetric
discretization is used but the transverse temperature gradients are
limited \citep{sharma07}. This method is almost as fast as the
precedent one (three time faster that the MHD solver) but the limiter
increases a lot the perpendicular diffusion
($\kappa_{num}/\kappa_{\|}=7 \times 10^{-4}$). The second method, from
this article, is based on a physically motivated discretization along
the magnetic flux tube. The accuracy then turns out to be better
($\kappa_{num}/\kappa_{\|}=2 \times 10^{-4}$) but it is a little bit
slower (although still one time and half faster than the MHD
solver).  In order to allow the reader to judge which problem size can be
realistically treated we give here an indication of the cpu time and
the memory consumption for a 3D run with $100^3$ grids on an AMD
Operon 1.8 GHz. A double time step consisting in two calls to the
transport subroutine and two calls to the MHD+cosmic rays routine
takes about 30s and uses about 500 MB of memory. We are currently working on a parallel version in order to make larger runs.

Since this method is a non-conservative method, we have also estimated the average fraction of energy lost per time step in the circular conduction test. These losses are limited to about $10^{-5}$, which is small considering that the magnetic field is strongly curved and the temperature falls by a factor of $10^4$ in few cells. Finally, it is worth noting that even higher level of anisotropy could be reached by implementing a higher order method. This could be easily done by interpolating the temperature field at higher order.

In the present and past section, we have shown that cosmic-ray tests
and passive conduction tests were succesfull. We now move to active
conduction tests which involve coupling between the MHD solver, the
cosmic-ray solver, as well as the anisotropic transport solver.

 \section{Buoyancy instabilities}
\label{sec:buoyancy} 

In this section, we focus on buoyancy instabilities in a stratified
atmosphere. We present here two applications of our code which serve
both as a test of our two-fluid code with anisotropic transport as
well as a physical study of the cosmic ray magnetothermal instability (CRMTI).

\subsection{Physical background \label{pcibackground}}

The cosmic ray magnetothermal instability (CRMTI) \citep{chandran06, dennis07} is a buoyancy instability that
is similar to the Parker instability \citep{parker66,shu74,ryu03} since it involves
magnetic fields and cosmic rays. 
However, in contrast to the Parker instability, the CRMTI involves anisotropic
thermal conduction and allows for a temperature gradient in the equilibrium, both
of which are relevant for understanding buoyancy instabilities in clusters
of galaxies. The CRMTI is very similar to the magnetothermal instability (MTI)
\citep{balbus00,balbus01,parrish05,parrish07}, except that it involves cosmic rays,
and, magnetic buoyancy if $\beta = 8\pi p/B^2$ is not large (as in the analysis of \citet{dennis07}). The stability
criterion is given by \citet{dennis07},
\begin{eqnarray}
n k_{B}dT/dz+dp_{cr}/dz+de_{mag}/dz>0 \label{criterion},
\end{eqnarray}
with $e_{mag}= B^2/8\pi$ the magnetic energy density. As discussed
by \citet{chandran05,chandran07}, the CRMTI may lead to convection
in galaxy cluster cores, since central AGN produce jets and centrally
concentrated cosmic rays. Such convection may play an important role in
transferring AGN power to the intracluster medium and helping
to solve the ``cooling flow problem.''

To understand the physics of this instability, let's take a simple
example where this instability applies. Consider a magnetized plasma
plus cosmic-ray stratified atmosphere initially at equilibrium with a
negligible gradient of magnetic field strength and temperature but a
negative gradient of cosmic-ray pressure ($dp_{cr}/dz<0$). Consider
also a vertical gravity along ``z'' $\bm{g}=-g \bm{e}_z$ and a
horizontal magnetic field along ``y'', $\bm{B}_0=B_0 \bm{e}_y$. If one
perturbs this medium by pushing upward a fluid parcel ($\delta v_z
>0$, with $\delta$ the difference between the perturbed state in the
bubble and the equilibrium state in the surrounding), the magnetic
field lines will be distorted ($\delta B_z \ne 0$). As a consequence
of the anisotropic transport, the cosmic ray pressure and plasma
temperature will be smoothed along the perturbed field lines, that is
to say the temperature and cosmic-ray pressure of the upwardly
displaced fluid parcel will be the same as in the initial
equilibrium. Therefore, the cosmic ray pressure in the upwardly
displaced fluid parcel will be larger than the one of the surrounding
medium ($\delta p_{cr}>0$) and the temperature will be almost the same
($\delta T \approx 0$). Requiring total pressure equilibration with
the surrounding medium, it implies that the parcel density is lower
than the one of the surrounding medium ($\delta \rho>0$) to compensate
for the high cosmic ray pressure. As a consequence, the parcel moves
upward faster, and the medium is convectively unstable. This
instability could even be amplified if the temperature gradient is
negative since in this case we would have $\delta T>0$. The latter
occurs in the magnetothermal instability (MTI), where $dp_{cr}/dz=0$
but $dT/dz<0$ \citep{balbus00}.

An important assumption in the above analysis is that $B_{z0} = 0$. If instead
$B_{z0} \neq 0$, then there is an equilibrium heat flux in the~$z$ direction, which
can further contribute to instabilities \citep{quataert08,parrish08}. However,
such ``heat-flux buoyancy instabilities'' are not considered further in this paper.

The linear analysis for the CRMTI has been done by \citet{chandran06}
and \citet{dennis07}. The dispersion relation is given by
\begin{eqnarray}
&&\omega^6-\omega^4 \left[ k^2 u^2+(k^2+k_y^2)v_A^2-g \frac{d\textrm{ln}\rho_0}{dz} \right] \\
&+&\omega^2 \left[k^2 k_y^2 v_A^2(2u^2+v_A^2)-(k_x^2+k_y^2)\left(g^2+(u^2+v_A^2)g \frac{d\textrm{ln}\rho_0}{dz} \right) \right] \nonumber\\
&+&k_y^2 v_A^2 \left[ -k^2 k_y^2 v_A^2 u^2+ (k_x^2+k_y^2) \left( g^2+u^2 g \frac{d\textrm{ln}\rho_0}{dz}\right) \right] =0,\nonumber
\end{eqnarray}
with $\omega$ the frequency, $\bm{k}=(k_x,k_y,k_z)$ the wave vector, $v_A^2=B_0^2/(4\pi\rho_0)$ the square of the Alfv\'en speed and
\begin{eqnarray}
u^2&=&\frac{p_0}{\rho_0} \frac{\gamma \omega+i \eta}{\omega+i\eta}+\frac{p_{cr0}}{\rho_0}\frac{\gamma_{cr} \omega}{\omega+i \eta},\\
\nu&=&k_y^2 D_{\|},\\
\eta&=&k_y^2 D_{\textrm{cond}}.
\end{eqnarray}

The eigenfunctions are given by,
\begin{eqnarray}
\delta \rho&=&-\frac{i\delta v_z}{\omega}\frac{d \rho_0}{dz}+\frac{\bm{k}.\delta \bm{v} \rho_0}{\omega}\\
\delta \bm{B}&=&-\frac{i\delta v_z}{\omega}\frac{d \bm{B}_0}{dz}-\frac{k_y B_0 \delta \bm{v}}{\omega}+\frac{\bm{k}.\delta\bm{v} \bm{B_0}}{\omega},\\
\delta p&=&-\frac{i\delta v_z}{\omega}\frac{d p_0}{dz}+\frac{\gamma \omega+ i \eta}{\omega+i \eta}\frac{\bm{k}.\delta \bm{v} p_0}{\omega},\\
\delta p_{cr}&=&-\frac{i\delta v_z}{\omega}\frac{d p_{cr0}}{dz}+\frac{\bm{k}.\delta \bm{v} \gamma_{cr} p_{cr0}}{\omega+i \nu},\\
\delta v_x&=&\frac{H.F-C.E}{A.E-H^2}\delta v_z,\\
\delta v_y&=&\frac{A.F-C.H}{H^2-A.E}\delta v_z,
\end{eqnarray}
with $A=-i[\omega^2-k_y^2 v_A^2-k_x^2(u^2+v_A^2)]$, $C=-k_x g+i k_x k_z (u^2+v_A^2)$, $E=-i(\omega^2-k_y^2 u^2)$, $F=i k_y k_z u^2-k_y g$ and $H=i k_x k_y u^2$. For a detailed discussion of this instability see \citet{chandran06, dennis07}. Let's now study this instability using our new code.

\subsection{An interesting limit: the magnetothermal instability (MTI)}

The easiest limit of this system is the adiabatic convective instability (no cosmic ray, no magnetic field, no conduction) which obeys the Schwarzchild stability criterion ($dS/dz>0$). As a preliminar test we have run a simulation and compared the evolution of the eigenfunction with the analytical one. The agreement between simulation and linear analysis is good. We do not present here the results since it has been abundantly studied in the literature. 

A more complex limit arises in dilute plasma when one takes into account anisotropic conduction. Indeed, in this case, the stability criterion becomes $dT/dz>0$ \citep{balbus00} and the medium could therefore be unstable even if the entropy gradient is positive. Since this instability has already been studied in \citet{parrish05}, our main purpose here is to test our code by comparing the numerical solution with the linear analysis. We therefore use a very similar test case to that studied by \citet{parrish05}. The main difference is that we trigger only two modes so that we are able to compute the exact linear solution (eigenfunctions).

Our initial conditions for the magnetothermal instability test are therefore a vertical equilibrium state for $-0.05<y<0.05$ and $-0.05<z<0.05$ (we use here the same axis as in \citet{chandran06} even if it's a 2D run):
\begin{eqnarray}
  \rho(z)&=&\rho_0 (1-\frac{z}{H_{\rho}}),\\
  p(z)   &=&p_0 (1-\frac{z}{H_{p}}),\\
  \bm{g}&=&-\frac{\bm{\nabla} p_{tot}(z)}{\rho(z)},\\
  \bm{B}&=& \sqrt{\frac{8\pi p_0}{\beta}}\bm{e}_y,\\
\end{eqnarray}
with $\rho_0=p_0=T_0=1$ in appropriate units and $\beta=2\times 10^6$ (to be in the high beta limit). By choosing $H_{\rho}=1.5$ and $H_{p}=1$, as in \citet{parrish05}, we ensure that the entropy gradient stays positive but the temperature gradient is negative. The conduction coefficient is $D_{\textrm{cond}}= 6.8\times 10^{-5}$. The box length $L=0.1$ satisfies the condition $L \le 0.1 min(H_{\rho},H_{p})$ which is essential to be able to apply local analysis. 

The boundary conditions need to conserve energy and also to not break artificially the initial equilibrium. We therefore choose periodic horizontal boundary conditions and reflective vertical boundary conditions. The implementation of reflective boundary conditions in presence of gravity turns out to be non trivial since the last cells are not in equilibrium if one uses standard reflective boundary conditions. In order to deal with this problem, we assign values to the ghost cells by assuming reflectional symmetry for scalars and reflectional antisymmetry for vectors, including gravity. We also remove the artificial diffusion of the last active cells that is normally added explicitly as part of the TVD method. In this way we could fullfill the two requirements: energy conservation and equilibrium.

We perturb the initial equilibrium with the superposition of two eigenfunctions with same $k_y=2 \pi /L$ but opposite $k_z$ ($k_z=+2 \pi /L$ and $k_z=-2 \pi /L$) so that the vertical velocity cancels along the horizontal boundaries. We take the amplitude of the velocity perturbation of each mode to be $\delta v_z/c_s=1.55\times 10^{-5}$ so that all the perturbed fields stay in the linear regime. Taking such a small value is very important since we will see that the relative amplitude of the magnetic field fluctuations are a factor $\approx c_s k/\sigma \approx 200$ times larger than the one of the velocity. To compute the eigenfunction we take the limit $p_{cr}=0$ and $D_{\|}=0$ of the eigenfunction of \citet{chandran06} described in the last subsection. We finally study the most unstable mode which is convective and exponentially growing. For this purpose, we run a 2D simulation with 200 by 200 grid points. The expected growth rate is $\sigma=0.22$. Note that to obtain the same results as \citet{parrish05} one has to trigger only one eigenfunction with $k_z=0$ but, in this case, boundary conditions then excite a lot of different and uncontrolled modes.

\begin{figure*}                                  \begin{tabular}{cc}
    \includegraphics[width=0.45\hsize]{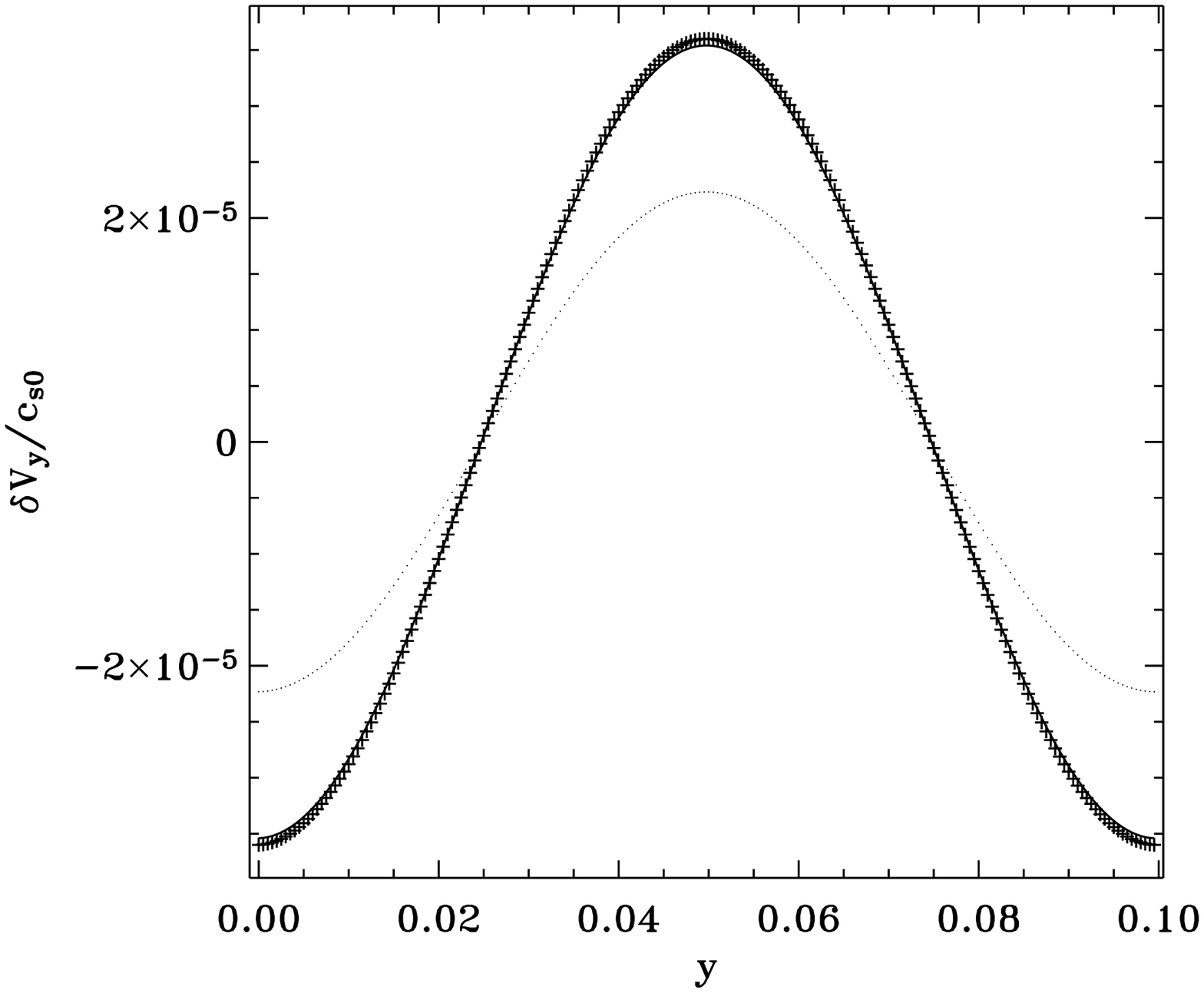}
    \includegraphics[width=0.45\hsize]{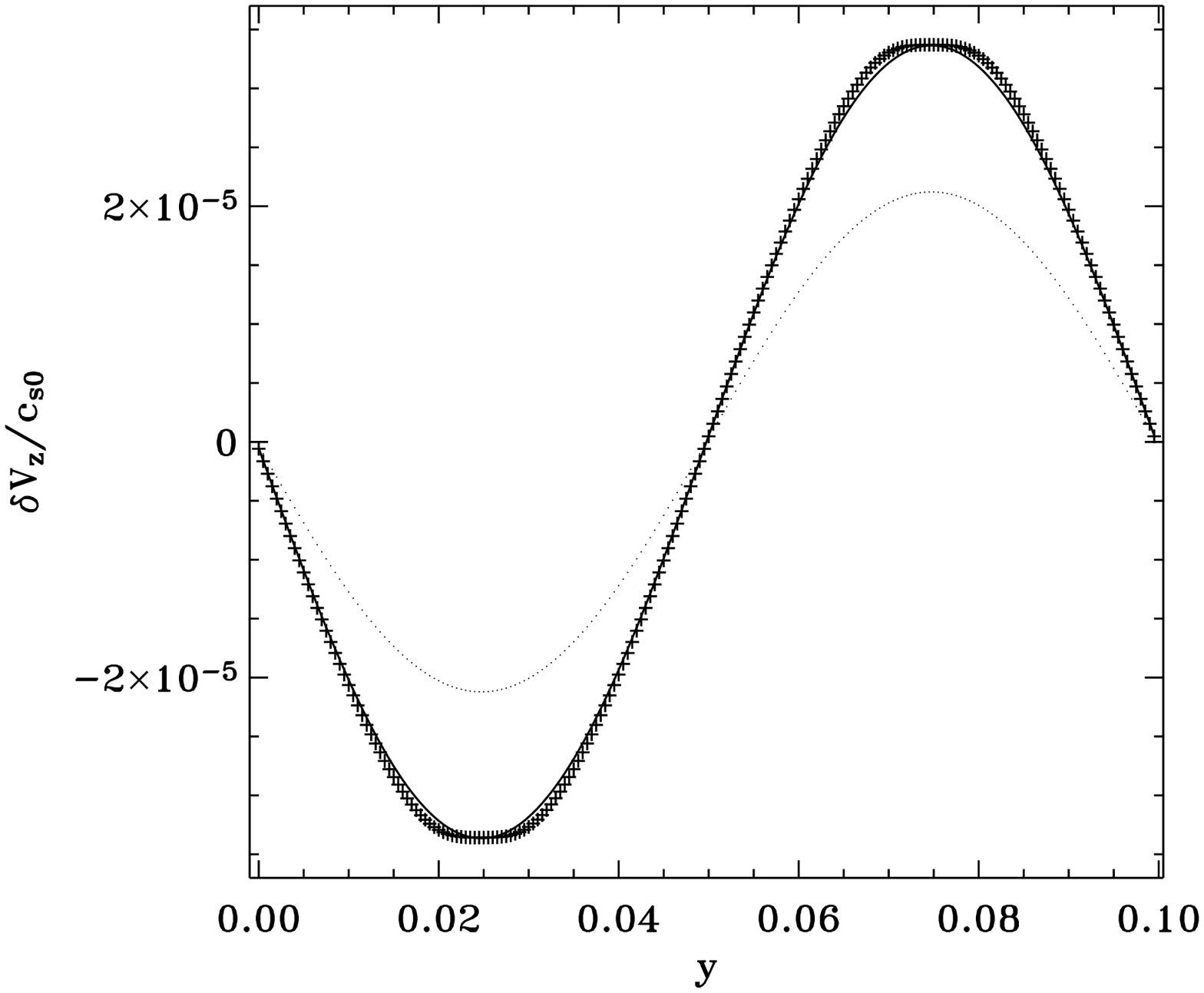}             \\
    \includegraphics[width=0.45\hsize]{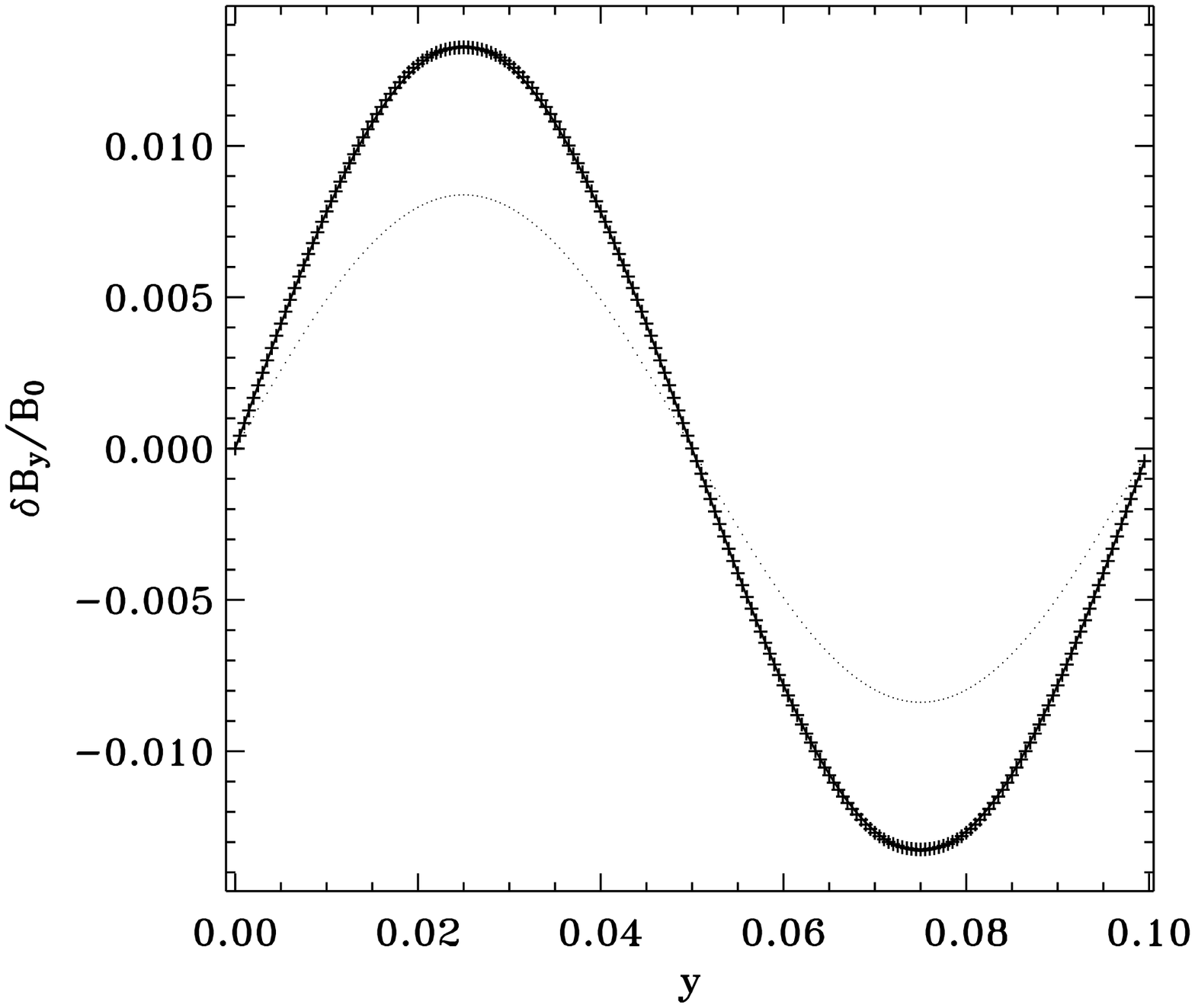}
    \includegraphics[width=0.45\hsize]{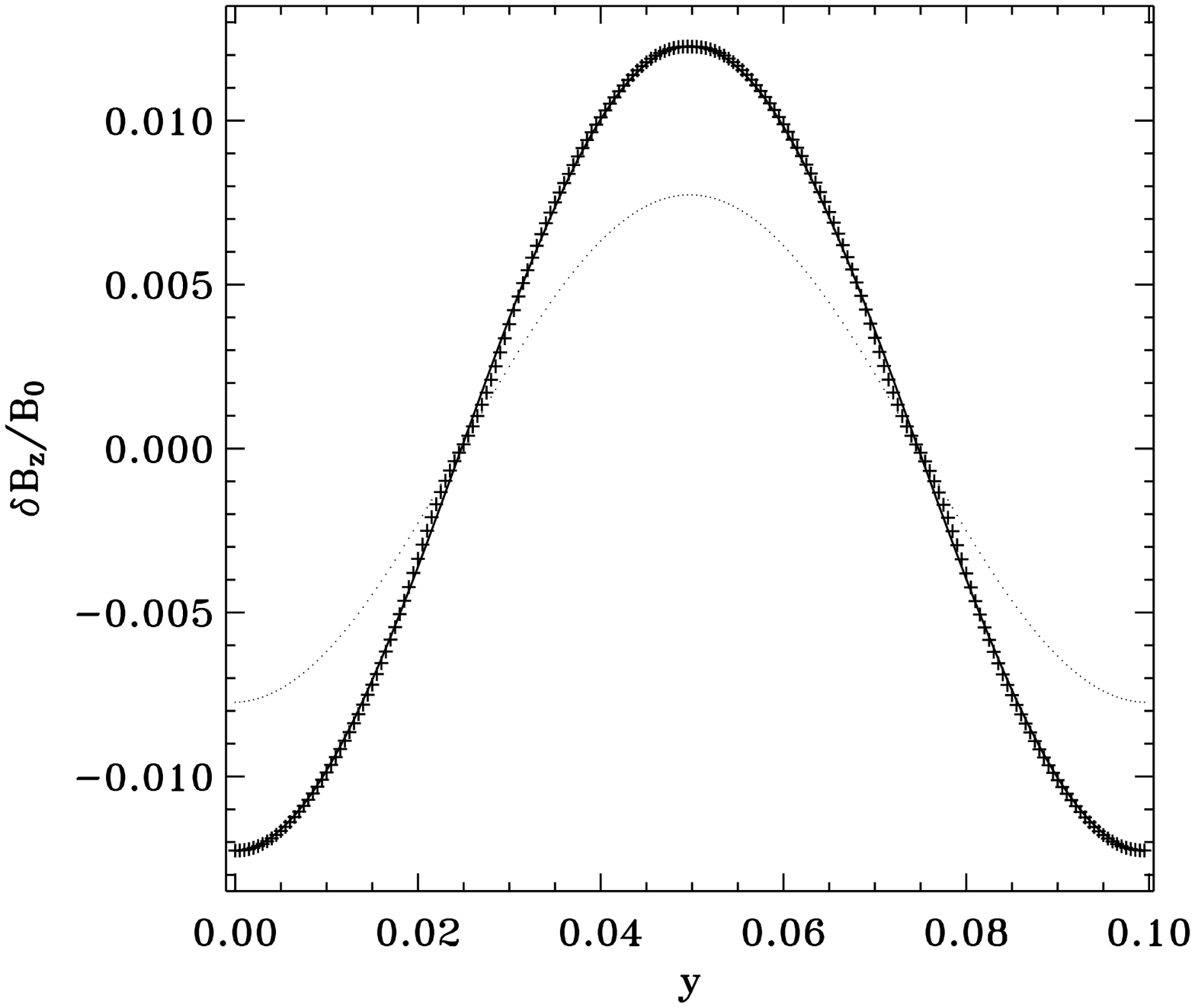}              \\
    \includegraphics[width=0.45\hsize]{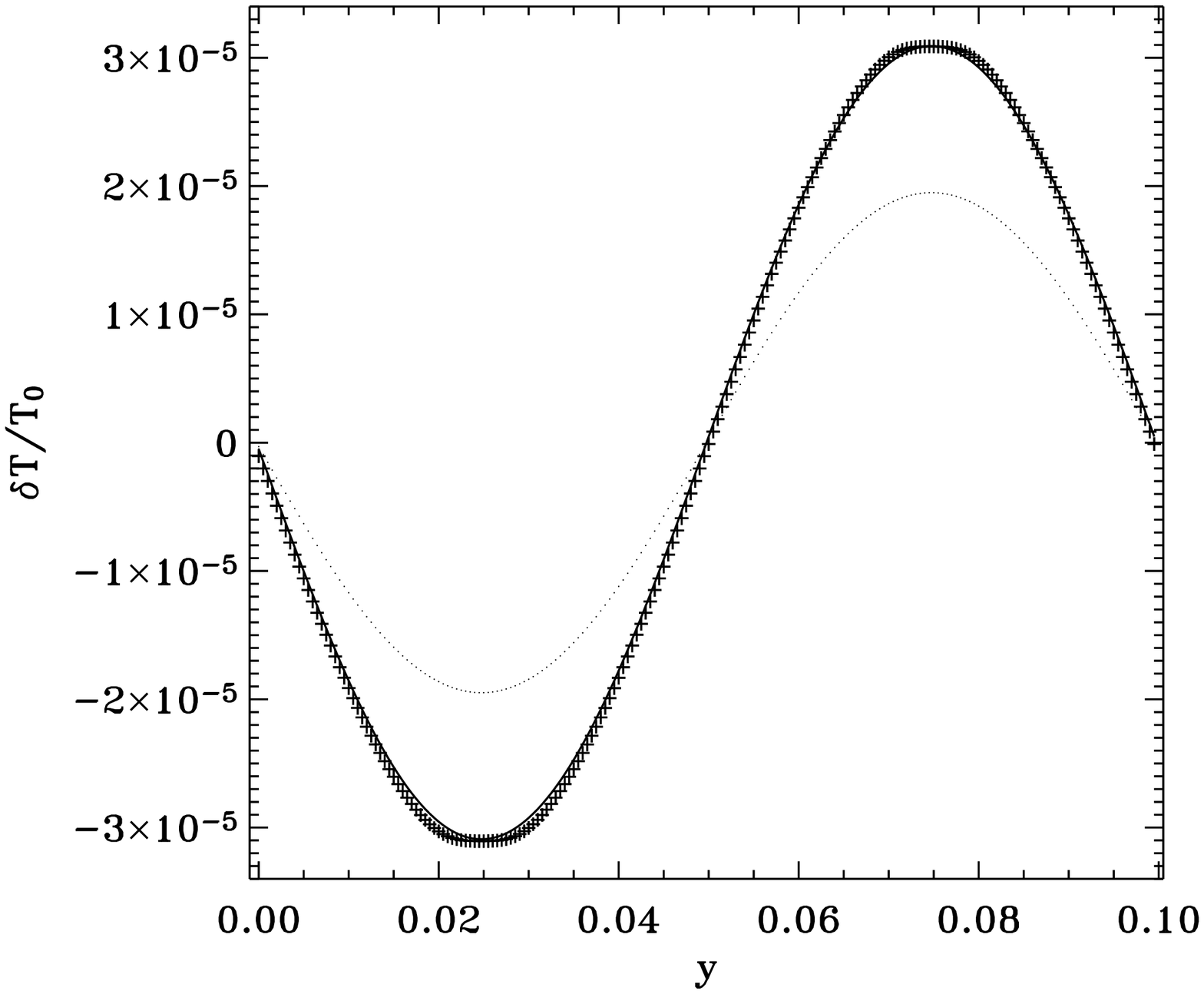}             
    \includegraphics[width=0.45\hsize]{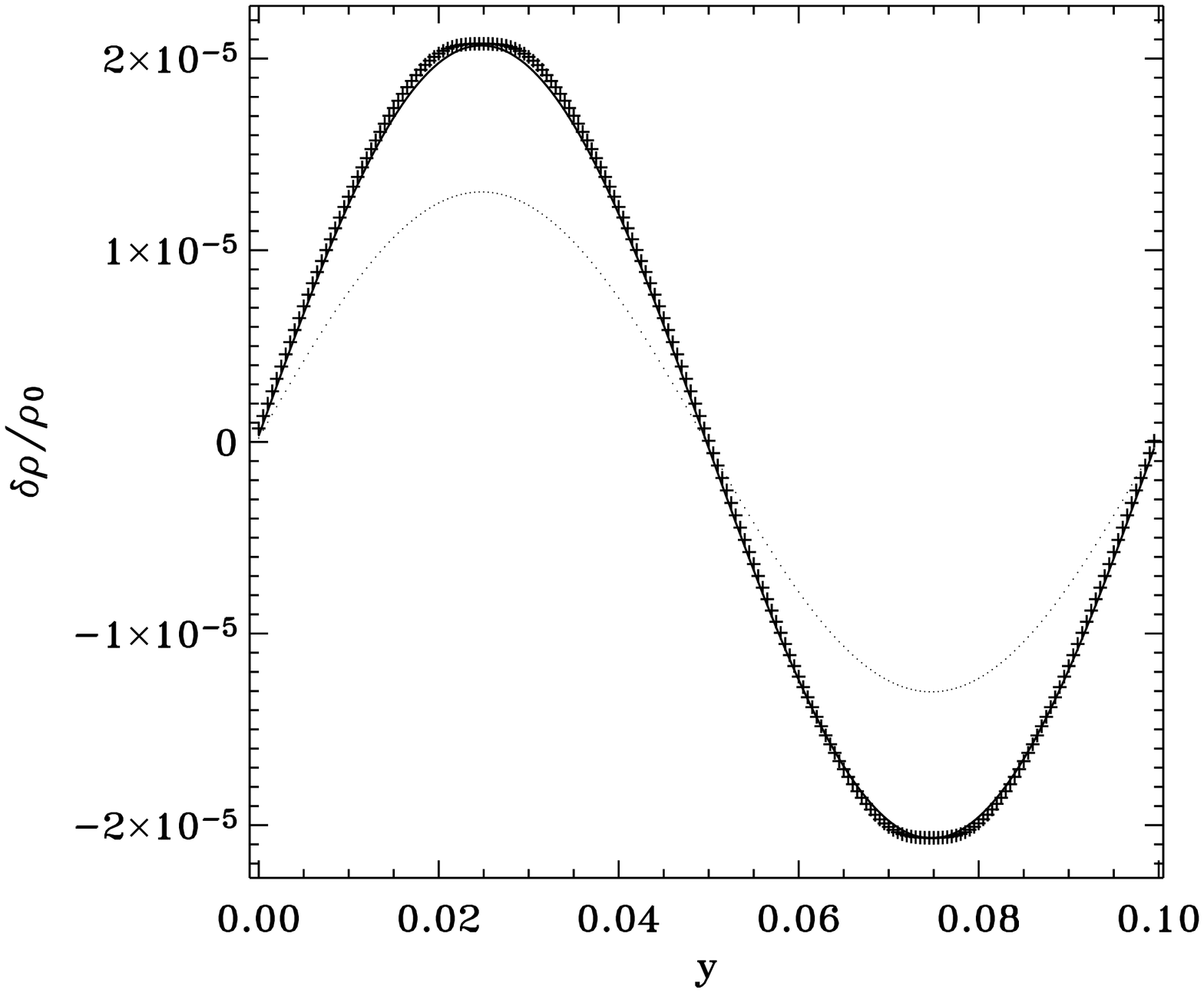}             
  \end{tabular}  
  \caption{Evolution of the eigenfunction in the code (plus signs) compared to the linear theory (solid line). The initial condition is the dotted line. The 6 graphs represent a slice in $z=0.0125$ for respectively $\delta v_y/c_s$, $\delta v_z/c_s$, $\delta B_y/B_0$, $\delta B_z/B_0$, $\delta T/T_0$ and $\delta \rho/\rho_0$ .\label{MTI}}
\end{figure*}

The results in Fig.\ref{MTI} show good agreement with the analytical result. We however note some slight deformations mainly due to the slope limiter as already mentioned in earlier sections. For example, one could clearly see the clipping of the maxima of $\delta v_z$. We have checked that without slope limiters these deformations do not appear. We therefore recover the results of \citet{parrish05}, concerning the growth rate. We have presented a test for active anisotropic conduction which turns out to be very strict and give no mercy to any approximations in any part of the scheme. We have noticed that a simple precision run or a too violent slope limiter quickly destroy the shape of the sines.


\subsection{Solution of the full system: the cosmic-ray magnetothermal instability (CRMTI)\label{pcisection}}

This code is especially dedicated to the study of this instability. This part serves as a cross-validation between our code and the analytical predictions of \citet{chandran06}. We use here the same kind of initial condition as in the study of the magnetothermal instability. The difference is that we have now cosmic rays and we therefore take,
\begin{eqnarray}
  p_{cr}(z)&=&p_{cr0} (1-\frac{z}{H_{cr}}).\\
\end{eqnarray}
Instead of choosing the same density and pressure gradients for the plasma as \citet{parrish05}, we prefer to study a different situation where the gradient of entropy and the gradient of temperature are positive. In this case, both the Schwarzchild stability criterion and the \citet{balbus00} stability criterion are satisfied. However, we choose a negative gradient of cosmic rays so that the atmosphere is convectively unstable according to Eq.\ref{criterion}. This kind of situation is expected in the center of cluster of galaxies \citep{chandran05,chandran06,dennis07,chandran07,rasera08}. Since we are mainly interested in the role of cosmic rays, we take a nul gradient of magnetic field so that the magnetic field doesn't modify the stability criterion \citep{dennis07}. 

Our initial conditions are inspired by the Perseus cluster at $50$~kpc \citep{chandran06}. Namely, we take $T_0=4$~keV, an electronic density $n_{e0}=0.02$~cm$^{-3}$, an ion density of $n_{i0}=0.9 n_e$, a mean molecular weigh of $\mu=0.6$, a cosmic ray pressure $p_{cr0}=p_{0}$ and a magnetic field of $B_0=1~\mu$G. For these values, the conduction coefficient is $D_{\textrm{cond}}=2.5\times10^{30}~$cm$^2/s$ and we choose the diffusion coefficient to be $D_{\|}=10^{29}~cm^2.s^{-1}$ (see Sect.\ref{sec:equations}). Concerning the gradient, we take a negative density gradient with $H_{\rho}=50$~kpc, a positive temperature gradient with $H_{T}=200$~kpc, and a negative cosmic ray gradient with $H_{cr}=50$~kpc. We finally use 100 by 100 grid points for a simulation box length of $L=10$~kpc. The amplitude of our two perturbations is $\delta v_z/c_s=1.44\times 10^{-5}$ and their wavelength is $10$~kpc. We use the same boundary conditions as before and the expected growth rate is $\sigma=9.5\times 10^{-3}$Myr$^{-1}$.

\begin{figure*}                                  \begin{tabular}{cc}
    \includegraphics[width=0.33\hsize]{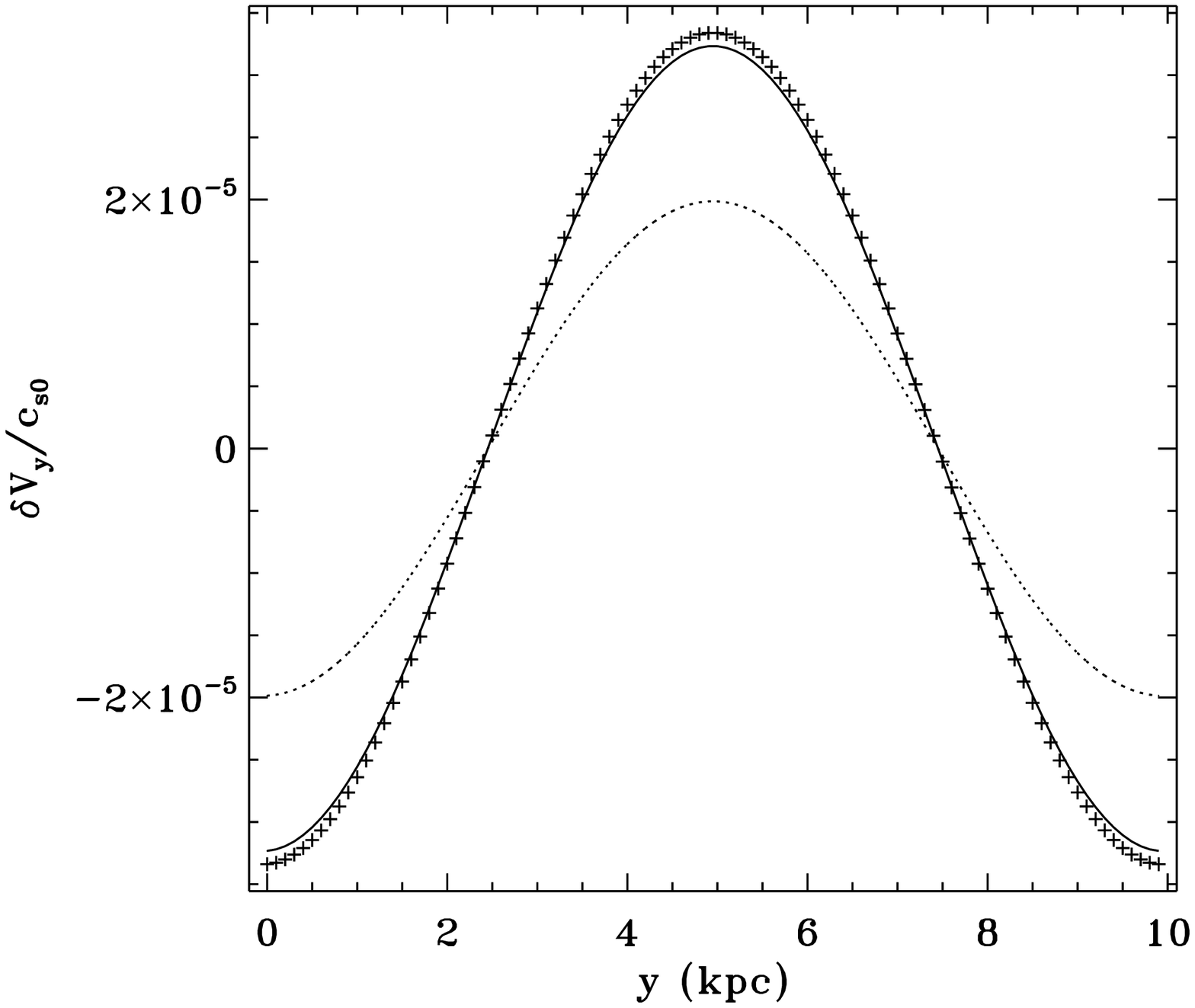}
    \includegraphics[width=0.33\hsize]{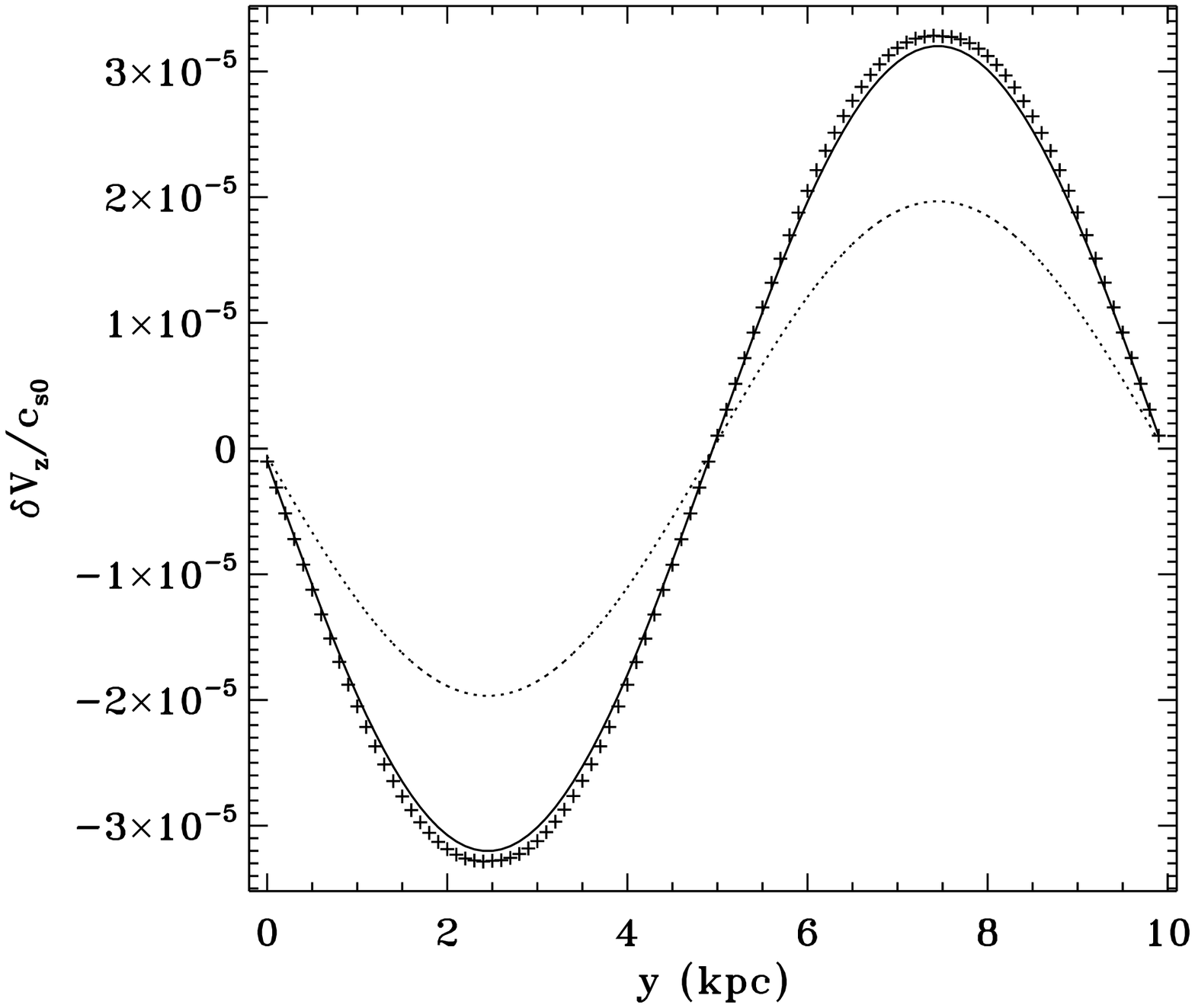}             \\
    \includegraphics[width=0.33\hsize]{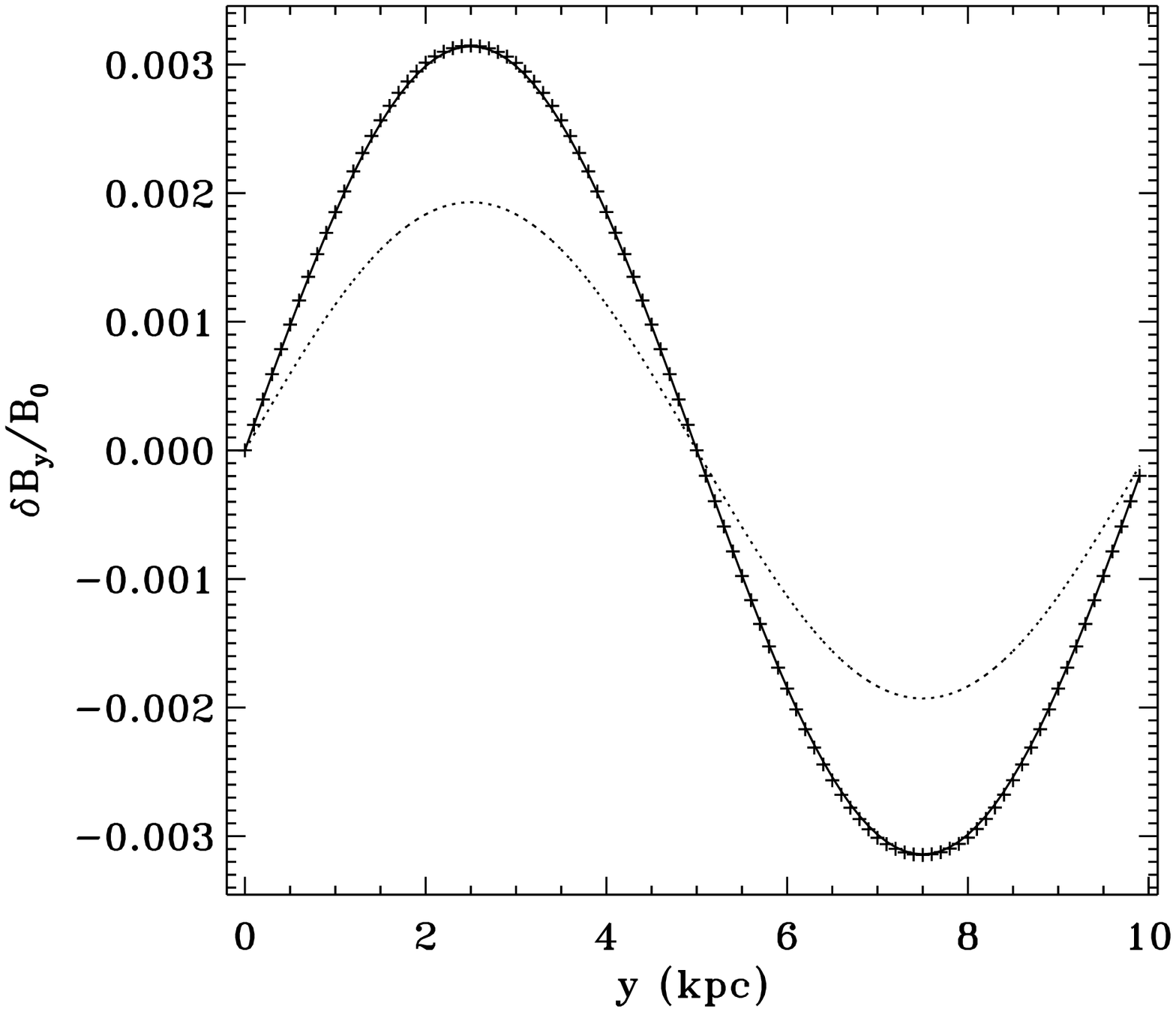}
    \includegraphics[width=0.33\hsize]{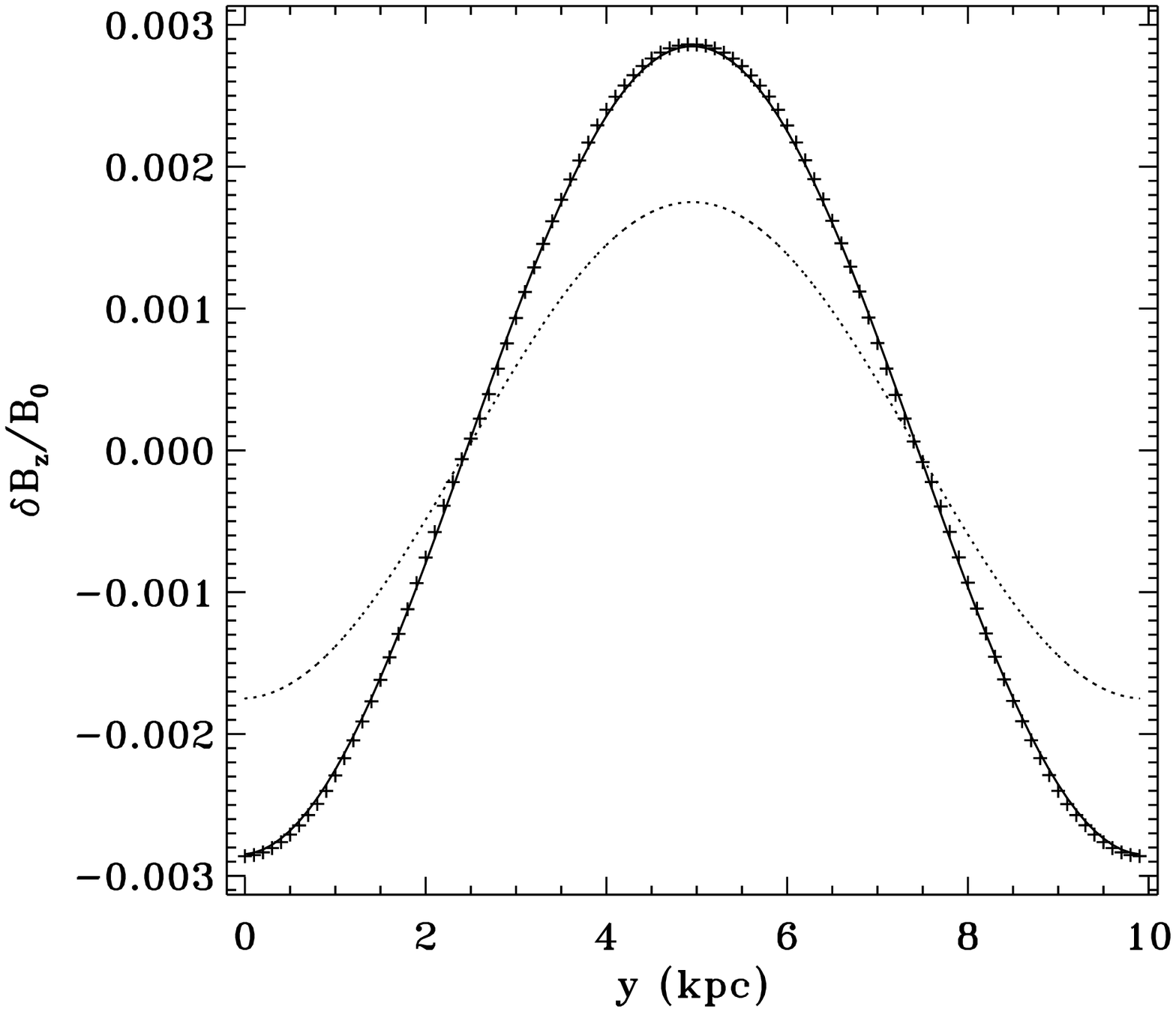}             \\
    \includegraphics[width=0.33\hsize]{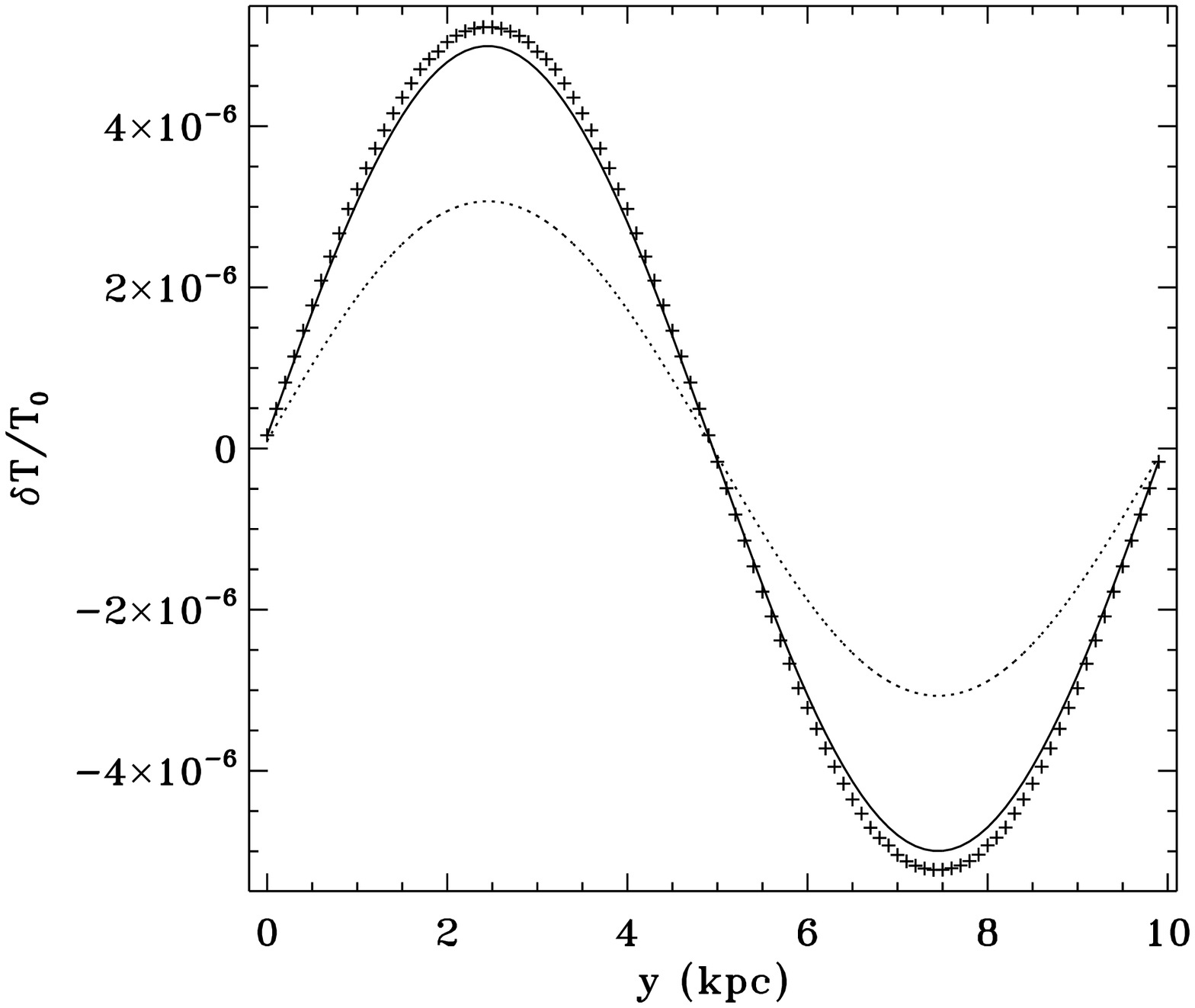}            
    \includegraphics[width=0.33\hsize]{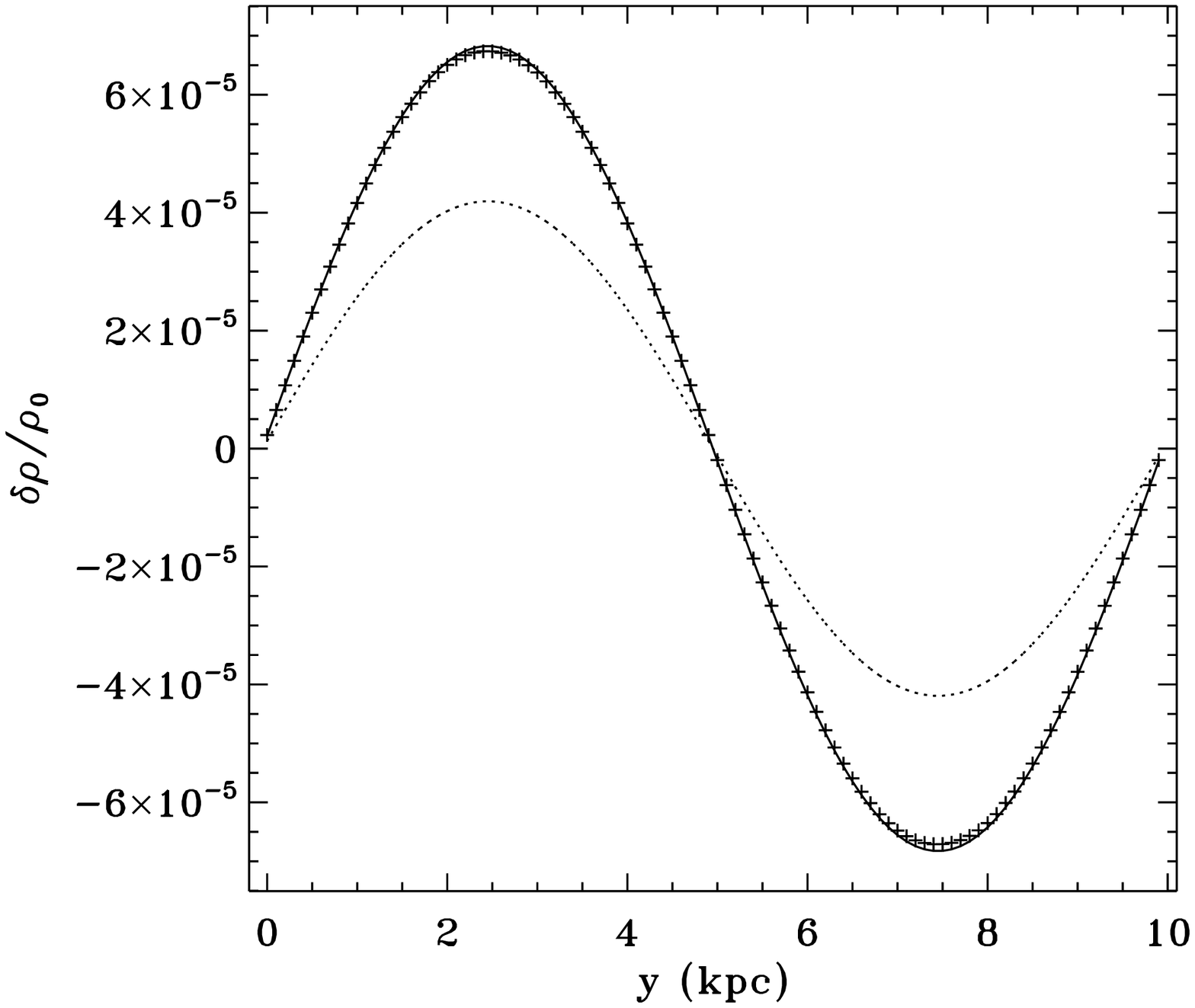}           \\
    \includegraphics[width=0.33\hsize]{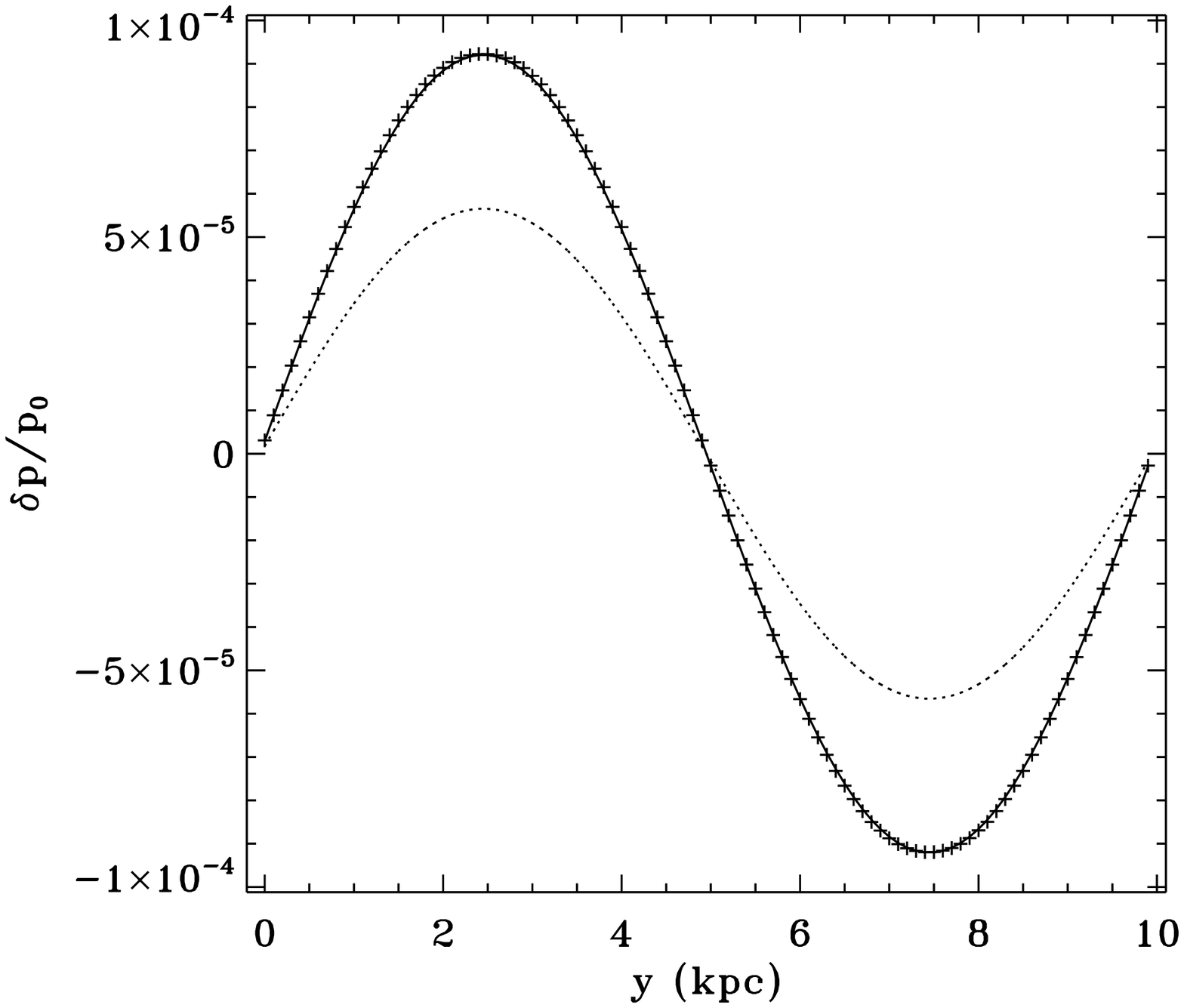}
    \includegraphics[width=0.33\hsize]{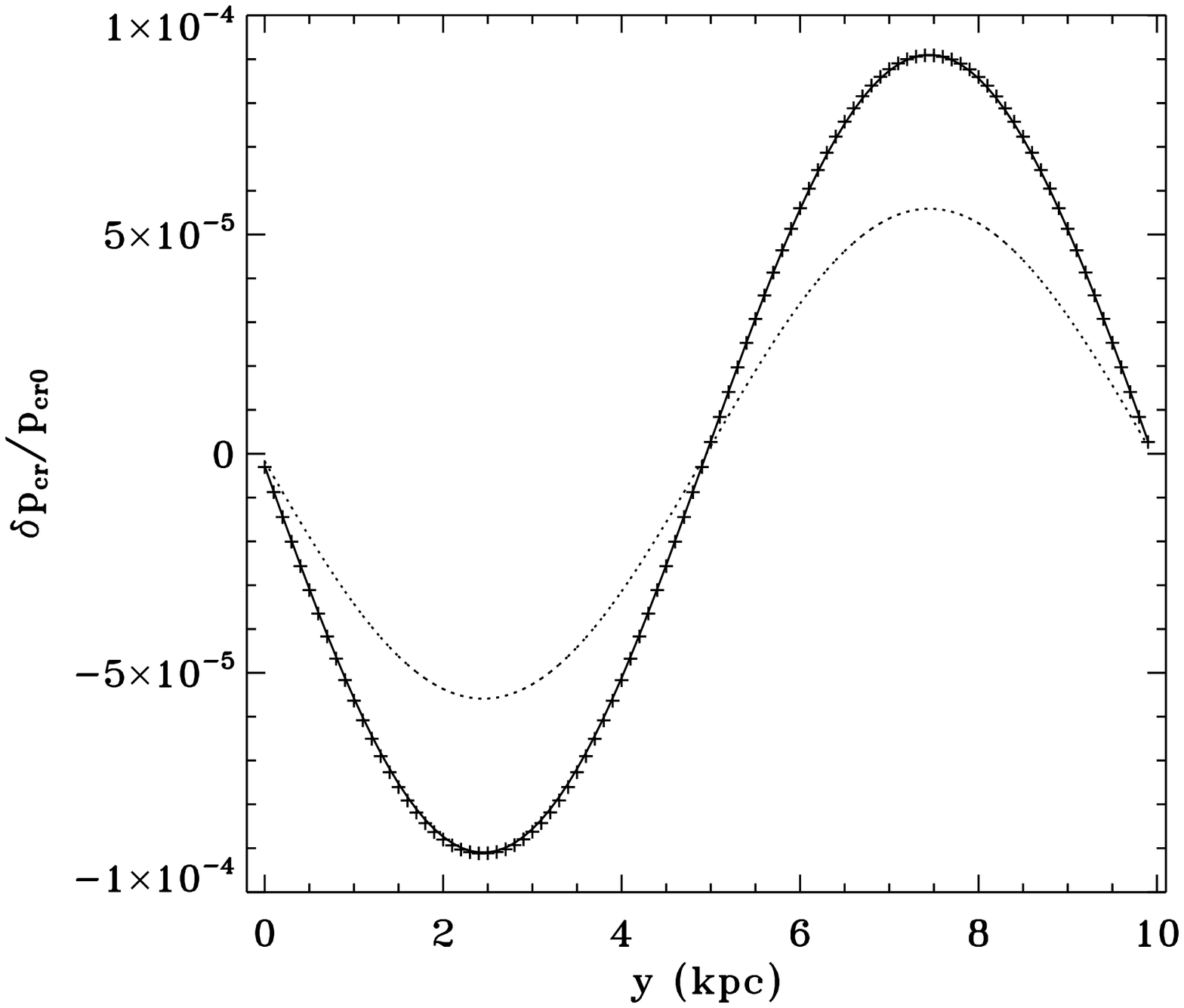}
  \end{tabular}  
  \caption{Evolution of the eigenfunction in the code (plus signs) compared to the linear theory (solid line, in most case covered by the plus signs). The initial condition is the dotted line. The 6 graphs represent a slice in $z=6.2$~kpc for respectively, $\delta v_y/c_s$, $\delta v_z/c_s$, $\delta B_y/B_0$, $\delta B_z/B_0$, $\delta T/T_0$, $\delta \rho/\rho_0$, $\delta p/p_0$ and $\delta p_{cr}/p_{cr0}$.     \label{PCI}}
\end{figure*}

Here again the results show good agreement with the linear theory (Fig.\ref{PCI}). The extrema of velocity are affected by the limiter. Little numerical errors appear on the temperature fluctuations because they have the smallest relative amplitude and are therefore the most sensitive to numerical approximations (such as the one induced by the slope limiter). The cosmic-ray pressure is computed with the same accuracy as the plasma pressure. This validates our overall implementation. 

On the physical point of view, this simulation illustrates the scenario presented in part \ref{pcibackground}. The phase of $\delta B_z$ is shifted from $\approx \pi/2$ compared to the phase of $\delta v_z$, since the field lines are distorted. The fluctuations of temperature $\delta T$ are very small. The perturbation $\delta p_{cr}$ is roughly in phase with $\delta v_z$ and is amplified, which in turn amplifies $\delta v_z$. Even though the entropy and temperature gradients are positive (and the magnetic field energy gradient is nul), the convective instability is growing on a timescale of $10^8$~yr which is of order of the cooling time near the center of cooling-flow clusters. This suggests that cosmic rays and anisotropic transport might play an important role in cluster of galaxies.

\section{Conclusion}

This article could be viewed as a test guide for those who want to implement cosmic-ray and anistropic transport routines, which are essential ingredients to simulate cooling-flow clusters. We indeed used many linear and non-linear tests for each physical ingredient as well as a new linear test for the full system. Our contribution could be divided into three parts. 

First, we showed that the TVD method of \citet{pen03} can be used to
evolve the cosmic rays and the plasma simultaneously. The shock tube
problem for a composite of plasma and cosmic rays is an example of the
successful implementation. Second, we insisted on the importance of
having a positive implementation of the anisotropic conduction in
order to ensure physical results even in the presence of sharp
gradients. We therefore presented a new flux-tube method which has two
important properties: positivity and accuracy. This is important for
clusters of galaxies since the conduction is highly anisotropic as
opposed to a very diffusive scheme. Moreover, the random magnetic
fields and potential large temperature and cosmic-ray-pressure
gradients in cluster cores may cause negative temperatures in a
non-positive scheme. Third, the linear regime of the cosmic ray
magnetothermal instability (CRMTI) provides a new, sensitive test, of
the overall implementation. The main interest is that this instability
has a broader range of applications since the criterion is $n
k_{B}dT/dz+dp_{cr}/dz+de_{mag}/dz>0$. One interesting future
application of this code concerns the cores of clusters of
galaxies. Indeed, in these regions, negative radial gradients of
cosmic-ray pressure may trigger convection. Moreover, possible large
gradients of cosmic-ray pressure near the edges of X-ray cavities
(cosmic-ray bubbles) require positive implementation of the
anisotropic transport.  It is worth noting that even though our
discussion has focused mainly on clusters of galaxies, the flux-tube
method that we have developed could be used to simulate a variety of
physical systems, including the interstellar medium, fusion plasmas,
and supernovae-remnants.  \acknowledgments

We are grateful to U.-L.~Pen, T.~Dennis, P.~Sharma, W.~Hammett, R.~Teyssier, I.~Parrish and J.~Stone for their valuable comments and suggestions.

\clearpage

\begin{table}[h!]
\begin{center}
\begin{tabular}{|c|c|c|c|c|c|}
\hline
\bf{Method}&positive&diffusion&speed2D&speed3D&losses\\
\hline
\bf{standard}&\bm{NO}&$10^{-4}$&$4.2$&4.6&$10^{-10}$\\
\hline
\bf{limited}&YES&$7 \times 10^{-4}$&$3.3$&3.8&$10^{-10}$\\
\hline
\bf{flux tube}&YES&$2 \times 10^{-4}$&$1.6$&1.0&$10^{-5}$\\
\hline
\end{tabular}
\caption{Summary of the three second order methods for anisotropic conduction or diffusion.\label{methods_anis}}
 \tablecomments{Columns indicate respectively: the method (standard asymmetric discretization, Van Leer limited or flux tube method), the positivity of the solution, the value of $\kappa_{num}/\kappa_{\|}$ in the Sovinec test, the speed relative to the MHD solver in 2D and 3D CRMTI instability test (see Sect.\ref{pcisection}) and the fraction of energy lost per time step at the end of the circular conduction test.}
\end{center}
\end{table}

\clearpage

\end{document}